\documentclass[useAMS,usenatbib]{mn2e}
\usepackage[dvips]{graphicx}
\usepackage{float}
\newcommand{\bz}{$\langle B_z \rangle$}
\newcommand{\nz}{$\langle N_z \rangle$}
\newcommand{\vsini}{$v \sin i$}
\newcommand{\kms}{km\,s$^{-1}$}
\newcommand{\halp}{H$\alpha$}

\newcommand{\mdot}{$\dot{M}$}
\newcommand{\vinf}{$v_\infty$}
\newcommand{\rsun}{$R_\odot$}
\newcommand{\msun}{$M_\odot$}
\newcommand{\lsun}{$L_\odot$}
\newcommand{\teff}{$T_{\rm eff}$}

\title[An observational evaluation of magnetic confinement in the winds of BA supergiants]{An observational evaluation of magnetic confinement in the winds of BA supergiants\thanks{Based on observations obtained at the Canada-France-Hawaii Telescope (CFHT) which is operated by the National Research Council of Canada, the Institut National des Sciences de l'Univers (INSU) of the Centre National de la Recherche Scientifique of France, and the University of Hawaii, and on observations obtained using the Narval spectropolarimeter at the Observatoire du Pic du Midi (France), which is operated by the INSU.}}

\author[M.~Shultz et al.]
{\parbox{\textwidth}{M.~Shultz,$^{1,2,3}$\thanks{E-mail: \texttt{mshultz@astro.queensu.ca}}
G.~A.~Wade,$^{2,1}$
V.~Petit,$^{4}$
J.~Grunhut,$^{5}$
C.~Neiner,$^{6}$
D.~Hanes,$^{1}$ and
the MiMeS Collaboration}\vspace{0.4cm}\\
\parbox{\textwidth}{$^{1}$Department of Physics, Engineering Physics, and Astronomy, Queen's University, Kingston, ON, Canada\\
$^{2}$Department of Physics, Royal Military College of Canada, Kingston, ON, Canada\\
$^{3}$European Southern Observatory, Santiago, Chile\\
$^{4}$Dept. of Geology \& Astronomy, West Chester University, West Chester, PA 19383\\
$^{5}$European Southern Observatory, Garsching, Germany\\
$^{6}$LESIA, UMR 8109 du CNRS, Observatoire de Paris, UPMC, Univ. Paris Diderot, 5 place Jules Janseen, 92195 Meudon Cedex, France\\
}}
\begin{document}

\date{Accepted 1988 December 15. Received 1988 December 14; in original form 1988 October 11}

\pagerange{\pageref{firstpage}--\pageref{lastpage}} \pubyear{2002}

\maketitle

\label{firstpage}

\begin{abstract}
Magnetic wind confinement has been proposed as one explanation for the complex wind structures of supergiant stars of spectral types B and A. Observational investigation of this hypothesis was undertaken using high-resolution ($\lambda/\Delta\lambda\sim~65,000$) circular polarization (Stokes $V$) spectra of six late B and early A type supergiants ($\beta$ Ori, B8Iae; 4 Lac, B9Iab; $\eta$ Leo, A0Ib; HR1040, A0Ib; $\alpha$ Cyg, A2Iae; $\nu$ Cep, A2Iab), obtained with the instruments ESPaDOnS and Narval at the Canada-France-Hawaii Telescope and the Bernard Lyot Telescope. Least Squares Deconvolution (LSD) analysis of the Stokes $V$ spectra of all stars yields no evidence of a magnetic field, with best longitudinal field 1$\sigma$ error bars ranging from $\sim$0.5 to $\sim$4.5 G for most stars. Spectrum synthesis analysis of the LSD profiles using Bayesian inference yields an upper limit with 95.4\% credibility on the polar strength of the (undetected) surface dipole fields of individual stars ranging from 3 to 30 G. These results strongly suggest that magnetic wind confinement due to organized dipolar magnetic fields is not the origin of the wind variability of BA supergiant stars. Upper limits for magnetic spots may also be inconsistent with magnetic wind confinement in the limit of large spot size and filling factor, depending on the adopted wind parameters. Therefore, if magnetic spots are responsible for the wind variability of BA supergiant stars, they likely occupy a small fraction of the photosphere.
\end{abstract}


\section{Introduction}

Early-type stars are amongst the brightest stars in spiral galaxies. These stars are massive ($M_*\ge 10$\msun), large ($R_*\sim$50--200 \rsun), and short-lived, with main-sequence lifetimes of 10--25 Myr. Despite their ephemeral existence and rarity, massive stars exert a disproportionate influence within the galactic ecology: in life their light creates ionized H{\sc ii} regions while their powerful winds play a decisive role in sculpting molecular clouds, acting to both trigger and quench star formation; in death, their supernovae synthesize (and distribute) all elements heavier than Fe (along with numerous lighter elements). 

Amongst the most luminous massive stars (up to several hundred thousand \lsun) are supergiants of spectral type late B through early A. The evolution of these BA supergiant (BA SG) stars is extremely rapid, making them particularly rare; the details of their evolution, influenced largely by their winds, have significant consequences for their mass and composition at the pre-supernova stage. BA SG stars are also of potential value to extragalactic astronomy, as they are luminous enough to be relatively easily observed in nearby galaxies, thus offering both a probe of metallicity and, via the wind-momentum--luminosity relationship, the promise of a distance indicator \citep{kudritzki1999}; the latter, obviously, must be calibrated by a proper understanding of the stellar wind. 

The closest BA SGs, such as Rigel ($\beta$ Ori, B8Iae) or Deneb ($\alpha$ Cyg, A2Iae), are amongst the brightest objects in the sky and have been extensively monitored, revealing a unique pattern of spectral variability dubbed `$\alpha$ Cyg variability' \citep{Lucy1976}. This is characterized by variability in both the radial velocities of metallic lines, and the detailed shapes of spectral line profiles, particularly the hydrogen Balmer $\alpha$ line (H$\alpha$), which evolves between P Cygni, inverse P Cygni, full absorption and full emission \citep{k1996a, k1996b}. $\alpha$ Cyg variability is also characterized by an absence of clear periodicities, evolving instead `semi-periodically': periodograms find no single dominant period, but rather a range of significant peaks with periods ranging from less than a day to months \citep{Lucy1976}. Periodogram peaks may also change from season to season, and typically differ between equivalent width and radial velocity measurements \citep{k1996a,k1996b,k1997,mark2008,rich2011,mora2012a}. 

The H$\alpha$ variability of $\alpha$ Cyg variables is impossible to reproduce using a simple spherically symmetric model of the stellar wind \citep{mark2008}, implying that the circumstellar environment of these stars is highly structured. Striking evidence for complex wind structures was found by \cite{k1996b}, who discovered $\beta$ Ori (B8Iae), HD 91619 (B7Ia) and HD 96919 (B9Ia) to exhibit a phenomenon known as High Velocity Absorption (HVA), in which the blue-shifted half of the \halp~line profile exhibits a sudden, large increase in absorption at high radial velocity. HVAs have since been noted in HD 199478 (B8Iae) and $\alpha$ Cyg (A2Iae) by \cite{mark2008} and \cite{rich2011}, respectively, and have been observed repeatedly in the case of $\beta$ Ori \citep{morr2008}.

\cite{is1997} suggested this puzzling behaviour to be due to magnetic fields: magnetic spots at the stellar surface might lead to a large loop of magnetically supported material within the wind environment. This is in rough analogy to the loops observed within the solar corona, although of much greater radial extent (a few stellar radii) and duration (HVAs can persist and evolve over months). With density and velocity fields substantially different from either the surrounding wind or the stellar surface, such loops might give rise to HVAs as they move in and out of view in corotation with the photosphere. 

When first posited, this hypothesis was considered plausible given existing polarimetric observations by \cite{sev1970}, from which a longitudinal field measurement of \bz$\sim$130 G suggested that $\beta$ Ori possessed a magnetic field of sufficient strength to make wind confinement feasible. A longitudinal magnetic field of $\sim2500\pm250$ G was also reported for $\nu$ Cep by \cite{scholzgerth1981}, who found \bz~to vary between null and maximum  over a timescale of years, based upon Zeeman spectrograms measured using an Abbe comparator \citep{gerth1977}. However, a more recent study of 7 A-type supergiants (including $\alpha$ Cyg, $\nu$ Cep, and 4 Lac) was performed by \cite{verd2003}, who used the MuSiCoS spectropolarimeter, and found no evidence of Zeeman signatures in the line profiles, and no 3$\sigma$ detections in the longitudinal field, with error bars ranging from 3--160 G. 

The most recent results concerning the magnetic properties of BA SGs have been presented by \cite{hub2012}, who observed HD 92207 (A0Iae) with the FORS2 instrument on the VLT, with 2 of the 3 observations yielding detections at greater than 3$\sigma$ in the longitudinal field, and \bz~measured at hundreds of G. However, some previous magnetic detections with FORS2 data, for $\beta$ Cep stars and slowly pulsating B (SPB) stars especially (e.g. \citealt{hub2007, hub2011a, hub2011b}), have not been confirmed using instrumentation with a higher spectral resolution \citep{silvester2009, shultz2012}, while reanalysis of published FORS measurements revealed that many reported magnetic detections may have been consequences of unaccounted-for instrumental effects or the specific algorithm used to reduce the data, rather than real photospheric magnetic fields \citep{bagn2012}. \cite{bagn2013} have re-analyzed the FORS data for HD92207, and have convincingly demonstrated that no detectable magnetic field is present in those data. Observation at high spectral resolution is thus necessary to consider the detection of magnetic fields definitive.

\begin{table*}
\centering
\caption[Summary of Stellar Parameters]{Summary of the sample stars' stellar, atmospheric, and wind parameters, with theoretical wind parameters calculated from the models described in Section 5. \halp~denotes the type of \halp~profiles historically observed for each star; HVAs notes whether High Velocity Absorption events have been witnessed in \halp; $B(\eta_*=1)$ gives the dipolar magnetic field necessary for minimal magnetic wind confinement (see Eq. \ref{etastar}). \\
\textbf{Reference key:} \\
a) \cite{firn2012}, b) \cite{prz2006}, c) \cite{mora2012a}, d) \cite{schil2008}, e) \cite{verd1999}, f) \cite{vink1999, vink2000, vink2001}, g) \cite{k1996a, k1996b}, h) \cite{rich2011}, i) \cite{ud2002}, j) \cite{mark2008a}, k) \cite{lyder2001}, l) \cite{mark2008}, m) \cite{BarlowCohen1977}, n) \cite{ches2010}, o) this work}
\begin{tabular}{l|llllll}
\hline
\hline
Parameter 	&	 HD 34085	&	 HD 212593	&	HD 87737	&	HD 21389	&	HD 197345	&	HD 207260 		\\
 	&	$\beta$ Ori, Rigel	&	 4 Lac	&	$\eta$ Leo	&	HR1040	&	$\alpha$ Cyg, Deneb	&	$\nu$ Cep 		\\
\hline														
$V$ (mag)$^{\rm a}$ & 0.138$\pm$0.032 & 4.569$\pm$0.018 & 3.486$\pm$0.053 & 4.54$^{\rm e}$ & 1.246$\pm$0.015 & 4.289$\pm$0.007 \\
Spectral Type$^{\rm a}$ 	&	B8 Ia	&	 B9 Ib	&	A0 Ib	&	A0 Iae$^{\rm e}$	&	A2 Ia	&	A2 Iab 		\\
log($L/L_\odot$) 	&	 5.08$\pm$0.08$^{\rm b,c}$	&	 4.79$\pm0.21^{\rm j}$	&	4.23$\pm$0.12$^{\rm b}$	&	4.74$^{\rm e}$	&	5.30$\pm$0.07$^{\rm d}$	&	4.71$^{\rm e}$		\\
Distance (kpc)  	&	0.24$\pm$0.05$^{\rm b,c}$	&	1.65$^{\rm j}$	&	0.63$\pm$0.09$^{\rm b}$	& 0.98$\pm$0.09$^{\rm k}$	&	0.80$\pm$0.07$^{\rm d}$	&	0.63$^{\rm e}$		\\
Radius ($R_\odot$) 	&	71$\pm$14$^{\rm b,c}$	&	59$\pm 16^{\rm j}$	&	47$\pm$8$^{\rm b}$	&	97$^{\rm e}$	&	203$\pm$17$^{\rm d}$	&	92$^{\rm e}$		\\
$T_{{\rm eff}}$ (K) 	&	 12100$\pm$150$^{\rm j}$	&	 11200$\pm$200$^{\rm j}$ 	&	9600$\pm$150$^{\rm j}$	&	9730$^{\rm e}$	&	8525$\pm$75$^{\rm d}$	&	9080$^{\rm e}$		\\
log\textit{g} (cgs) 	&	 1.75$\pm$0.10$^{\rm j}$	&	 2.1$\pm$0.1$^{\rm j}$	&	2.00$\pm$0.15$^{\rm j}$	&	1.75$^{\rm e}$	&	1.10$\pm$0.05$^{\rm d}$	&	1.67$^{\rm e}$		\\
$M_{{\rm evol}} (M_\odot)$ 	&	 21$\pm$3$^{\rm b,c}$	&	21$\pm 6^{\rm j}$	&	10$\pm$1$^{\rm b}$	&	19.3$^{\rm e}$	&	18$\pm$2$^{\rm d}$	&	14.7$^{\rm e}$		\\
\hline														
\textit{v}sin\textit{i} (km/s) 	&	 25$\pm$3$^{\rm j}$	&	6$\pm 3^{\rm p}$	&	0$\pm$3$^{\rm j}$	&	25$\pm5^{\rm p}$	&	20$\pm$2$^{\rm d}$	&	15$\pm5^{\rm p}$		\\
P$_{\rm max}$ (d) 	&	143$^{+52}_{-40}$	&	497$^{+767}_{-256}$	&	792$^{+135}_{-135}$	&	196$_{-33}^{+49}$	&	513$^{+ 106}_{-85}$	&	310$_{-77}^{+155}$		\\
P$_{{\rm min}}$ (d) 	&	28$_{-10}^{+12}$	&	20$^{+15}_{-9}$	&	22$_{-7}^{+7}$	&	46	&	146$^{+24}_{-22}$	&	26		\\
\hline														
\textit{v}$_\infty$ (km/s) 	&	207	&	235	&	195	&	202	&	109	&	166		\\
$\dot{M} (M_\odot/{\rm yr})^{\rm f}$	&	 7.77$\times10^{-6}$ 	&	 1.64$\times10^{-6}$ 	&	 2.17$\times10^{-7}$ 	&	 1.22$\times10^{-6}$ 	&	 2.01$\times10^{-5}$ 	&	 1.40$\times10^{-6}$ 		\\
\textit{v}$_\infty$ (km/s) (measured) & 129$^{\rm l}$ & 350$^{\rm j}$ & -- & -- & 240$\pm$25$^{\rm d}$ & -- \\
$\dot{M} (M_\odot/{\rm yr})$ (measured) & 1.1$\times10^{-7, ~\rm m, n}$ & 9.1$\times 10^{-8, ~\rm j}$ & 4.7$\times 10^{-8, ~\rm m}$ & 4.2$\times 10^{-7, ~\rm n}$ & 3.1$\times 10^{-7, ~\rm d, n}$ & -- \\

H$\alpha$	&	variable$^{\rm g,h}$	&	full absorption$^{\rm j}$	&	full absorption	&	Double-peaked$^{\rm h}$	&	P Cygni$^{\rm g,h}$	&	P Cygni$^{\rm h}$		\\
HVAs?	&	many times$^{\rm g,h}$	&	--	&	--	&	yes$^{\rm h}$	&	yes$^{\rm h}$	&	yes$^{\rm h}$		\\
\hline
$B_{\rm d}(\eta_*=1){\rm ~(G)}^{\rm o}$  	&	40	&	24	&	10	&	12	&	17	&	12	\\
$B_{\rm d}(\eta_*=25){\rm ~(G)}^{\rm o}$  	&	203	&	119	&	50	&	58	&	83	&	60	\\
$B_{\rm sp}(\eta_*=200){\rm ~(G)}^{\rm o}$  	&	575	&	338	&	141	&	164	&	240	&	168	\\
\hline
$B_{\rm d}(\eta_*=1){\rm ~(G)}^{\rm o}$  	&	4	&	7	&	5	&	7	&	3	&	--	\\
$B_{\rm d}(\eta_*=25){\rm ~(G)}^{\rm o}$  	&	19	&	34	&	23	&	34	&	15	&	--	\\
$B_{\rm sp}(\eta_*=200){\rm ~(G)}^{\rm o}$  	&	54	&	98	&	66	&	96	&	43	&	--	\\
\hline
\hline
\end{tabular}
\label{sad-tab-2}
\end{table*}

This paper presents the results of observations of 6 BA SG stars obtained using high-dispersion spectropolarimeters, the most sensitive observations yet obtained for this class of stars. In Section 2, we describe the acquisition and reduction of the spectropolarimetric data-set. In Section 3, the wind behaviour during the period of observation is characterized.  In Section 4 we present the magnetic analysis of the data, and the resulting detection probabilities and longitudinal field measurements, together with constraints on the field strength obtained by means of spectral modeling interpreted using Bayesian statistical concepts. In Section 5 the implications of these results are interpreted in the context of magnetic wind confinement models, for both the case of a dipolar magnetic field, and small-scale magnetic spots.  

\section{Observations}

In the context of the Magnetism in Massive Stars (MiMeS) Large Programs, circular polarization (Stokes $V$) observations were collected with the cross-dispersed \'echelle spectropolarimeters ESPaDOnS (at the 3.6 m Canada-France-Hawaii Telescope (CFHT) on Mauna Kea) and its twin Narval (installed at the 2 m Telescope Bernard Lyot (TBL) at the Pic du Midi in the French Pyr\'en\'ees). Each instrument covers the spectral range from 369--1048 nm over 40 mostly overlapping spectral orders, with a spectral resolution of $\lambda/\Delta\lambda\sim$ 65,000 at 500 nm. The quality of the observations is in general excellent, with the peak signal-to-noise ratio per 1.8 \kms~pixel ranging from 450 to about 1300, and mean signal-to-noise ratios (SNRs) of 883 in ESPaDOnS spectra and 698 in Narval spectra. This data-set represents the highest-quality spectropolarimetry yet collected for these stars.

Each spectropolarimetric sequence consisted of four individual subexposures taken in different orientations of the half-wave Fresnel rhombs. From each set of four subexposures we derived Stokes $I$ and Stokes $V$ spectra following the double-ratio procedure described by \cite{d1997}, ensuring in particular that all spurious signatures of stellar or instrumental origin are removed at first order. Diagnostic null polarization spectra (labeled $N$) were calculated by combining the four subexposures in such a way that stellar polarization cancels out, allowing verification that no spurious signals are present in the data (see \cite{d1997} for more details on the definition of $N$). Frames were processed using the automated reduction package Libre-ESpRIT \citep{d1997}. The spectra were normalized by applying a manually-monitored order-by-order normalization employing $\sigma$-clipping and polynomial fits. 

In the cases of $\beta$ Ori and $\alpha$ Cyg, their brightnesses resulted in subexposure times (a few seconds) that were much shorter than the 25--40 s CCD read-out time (see Table \ref{btab-mean}). This makes spectropolarimetric monitoring (in the case of $\beta$ Ori, for which 65 observations were collected) or single-night deep observations (in the case of $\alpha$ Cyg, for which 100 observations were collected, 80 on two subsequent nights) a feasible investment. The remaining stars are much dimmer (see Table \ref{sad-tab-2}), meaning longer subexposure times were required, precluding the possibility of collecting more than a few observations in each case: 15 observations of $\eta$ Leo, 17 of $\nu$ Cep, 11 of 4 Lac, and 3 of HR 1040. Observation dates are given in Table \ref{btab-mean}, and span ranges from a few years ($\alpha$ Cyg and $\nu$ Cep) to one week (4 Lac).

Spectropolarimetric monitoring of a BA SG star presents special challenges as the rotational periods of these stars remain generally undetermined. However, minimum and maximum rotational periods (see Table \ref{sad-tab-2}) can be calculated based upon their projected rotational velocities $v\sin{i}$, surface gravities $\log{g}$, stellar radii $R_*$, and masses $M_*$. The stellar radius and \vsini~give the maximum period $P_{rot}/\sin{i}$, while the minimum period is established via the rotational breakup velocity $v_{\rm break} = \sqrt{GM_*/R_*}$ (assuming solid-body rotation). For $\beta$ Ori, $\eta$ Leo, and $\alpha$ Cyg, literature values were used for \vsini; for 4 Lac, HR 1040, and $\nu$ Cep, published rotational velocities were not compatible with observed line broadening, and \vsini~was estimated by fitting synthetic mean line profiles to the mean line profiles used in the magnetic analysis described in section 4. In calculating $P_{\rm min}$, the equatorial radius $R_{{\rm eq}}$ was used in place of $R_*$, since near critical velocity oblateness will mean that $R_{{\rm eq}} \simeq 1.5 R_{{\rm pol}}$, where $R_{{\rm pol}}$ is the polar radius. For $\eta$ Leo, only an upper limit is available for \vsini; the range of rotational periods is calculated using this upper limit. The effect is to increase $P_{\rm min}$ slightly, due to the increased radius, decreased surface gravity, and hence decreased $v_{\rm break}$.

In the cases of $\beta$ Ori and $\alpha$ Cyg, $R_*$ is derived from the Hipparcos parallax distances and interferometric angular radii \citep{perry1997, auf2002, auf2008}; for other stars $R_*$ is determined from stellar evolution models \citep{prz2006, prz2010, mark2008a, firn2012}. Observations of $\beta$ Ori were timed so as to span at least one rotational period but also to achieve good sampling at time-scales closer to $P_{min}$, with the most densely time-sampled period coinciding with a 28-day photometric observing campaign conducted with the Microvariability and Oscillations in STars (MOST) space telescope, intensive ground-based spectroscopic monitoring, and interferometry \citep{mora2012a,mora2012b,k2012}. 

\section{Wind Variability}

\begin{figure*}
\centering
\begin{tabular}{ccc}
\includegraphics[width=2.25in]{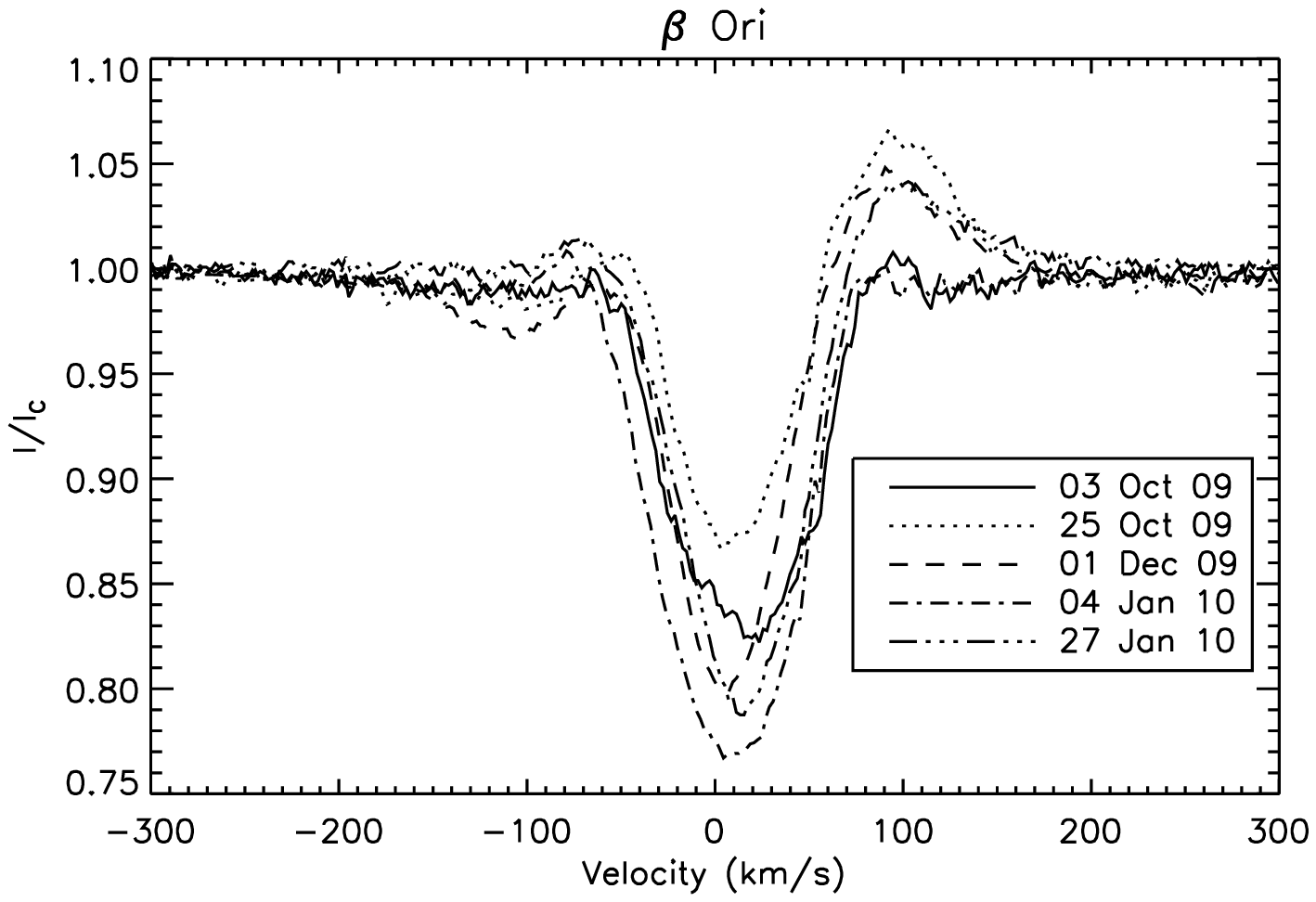} &
\includegraphics[width=2.25in]{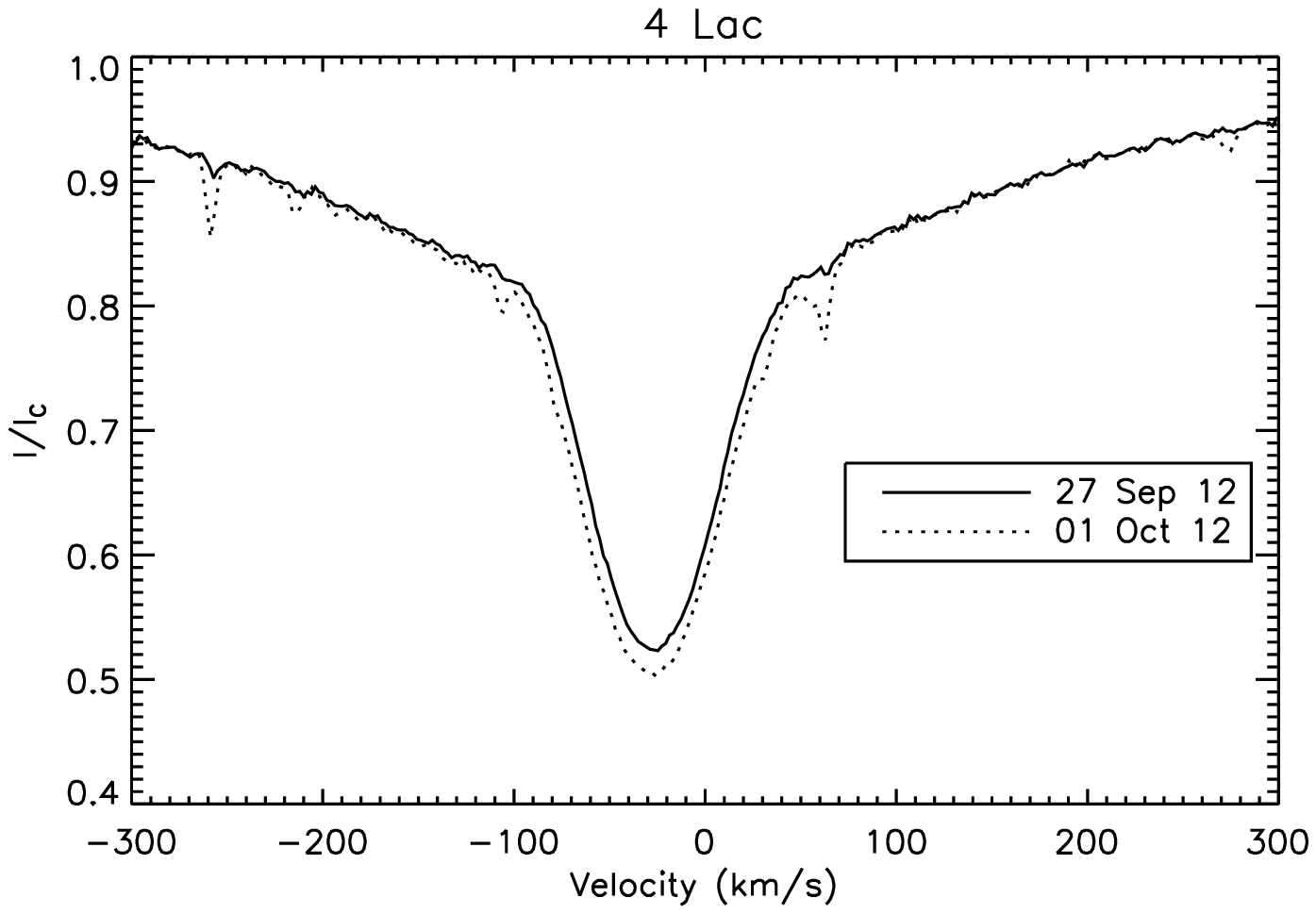} &
\includegraphics[width=2.25in]{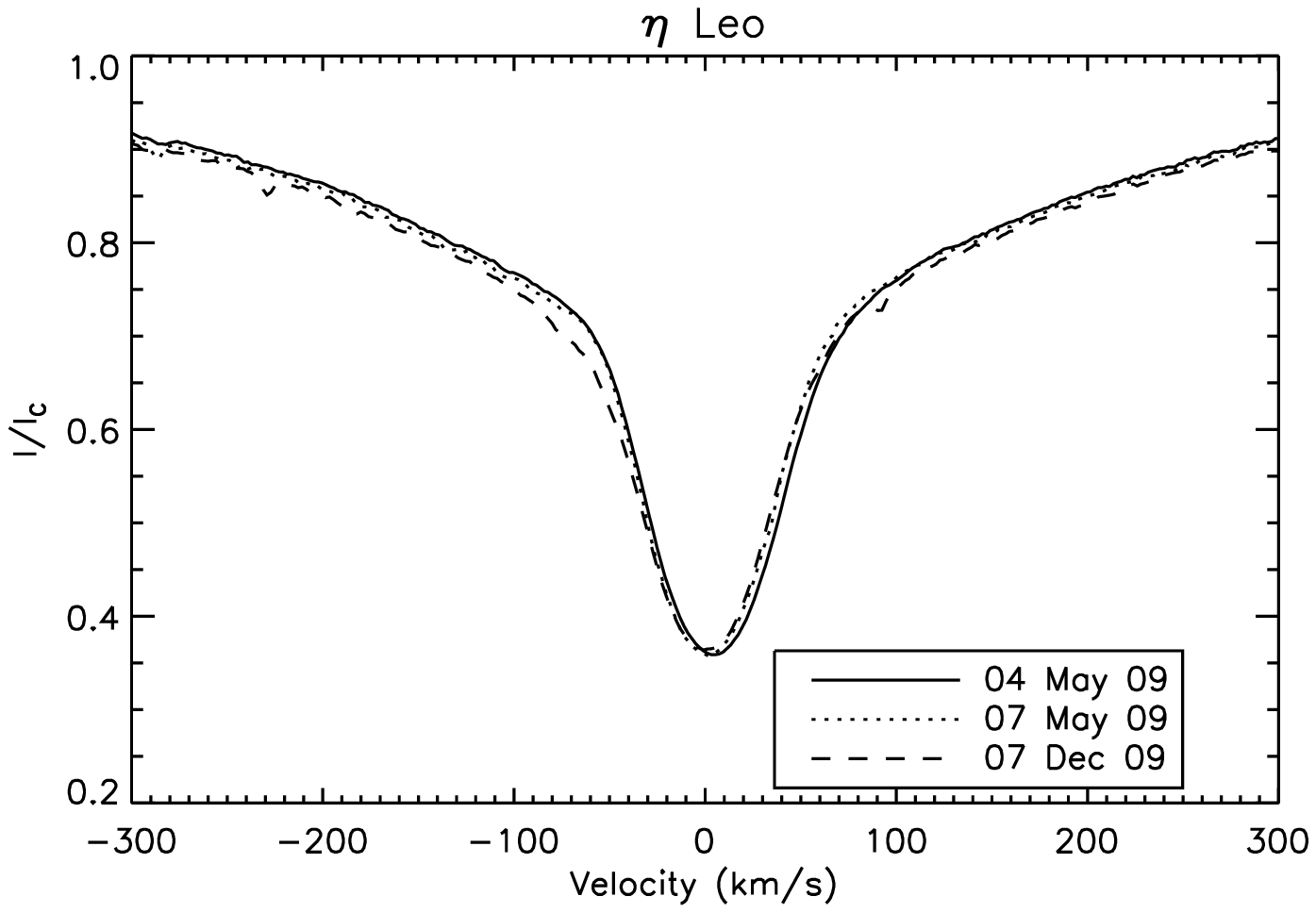}  \\
\includegraphics[width=2.25in]{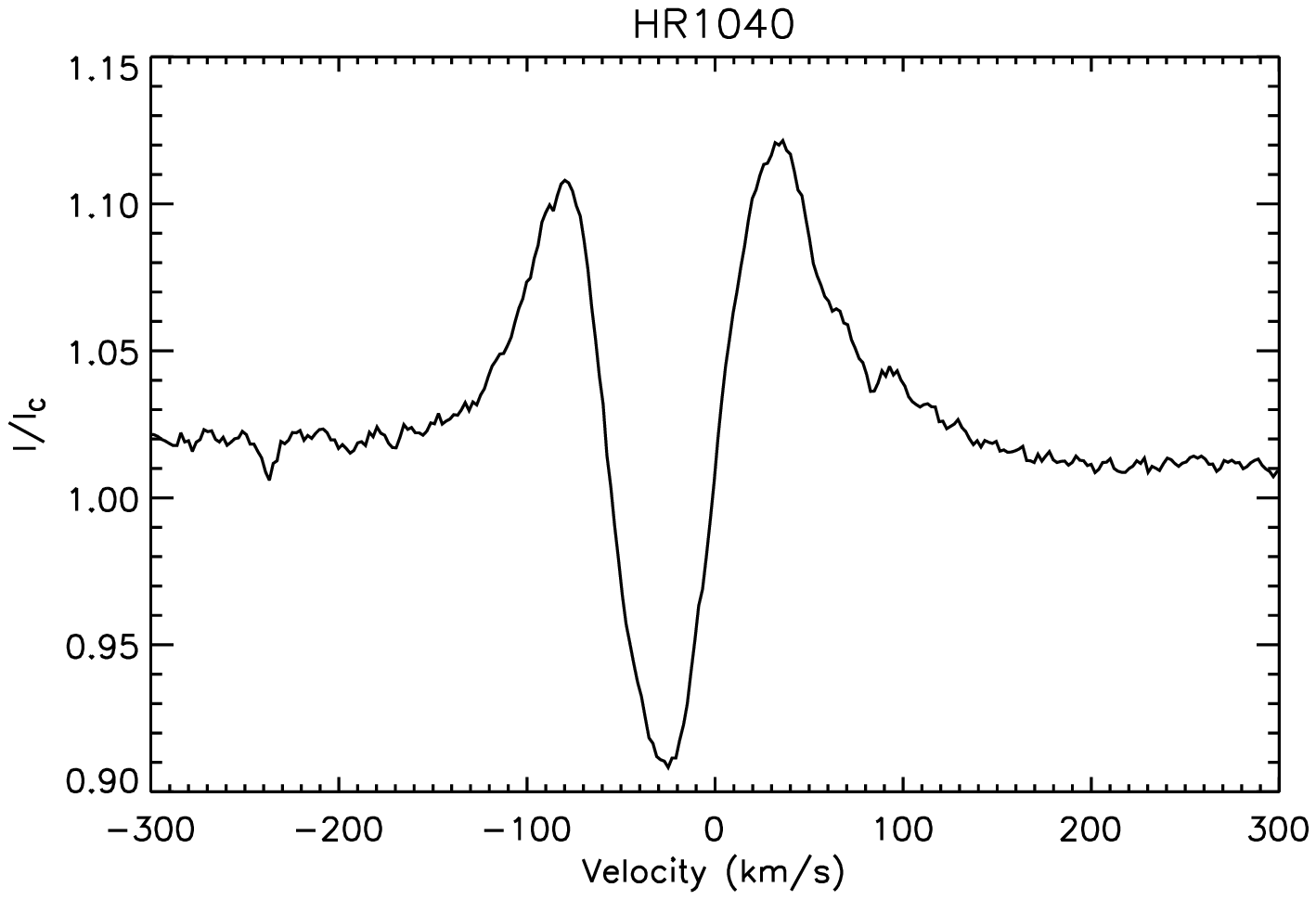} &
\includegraphics[width=2.25in]{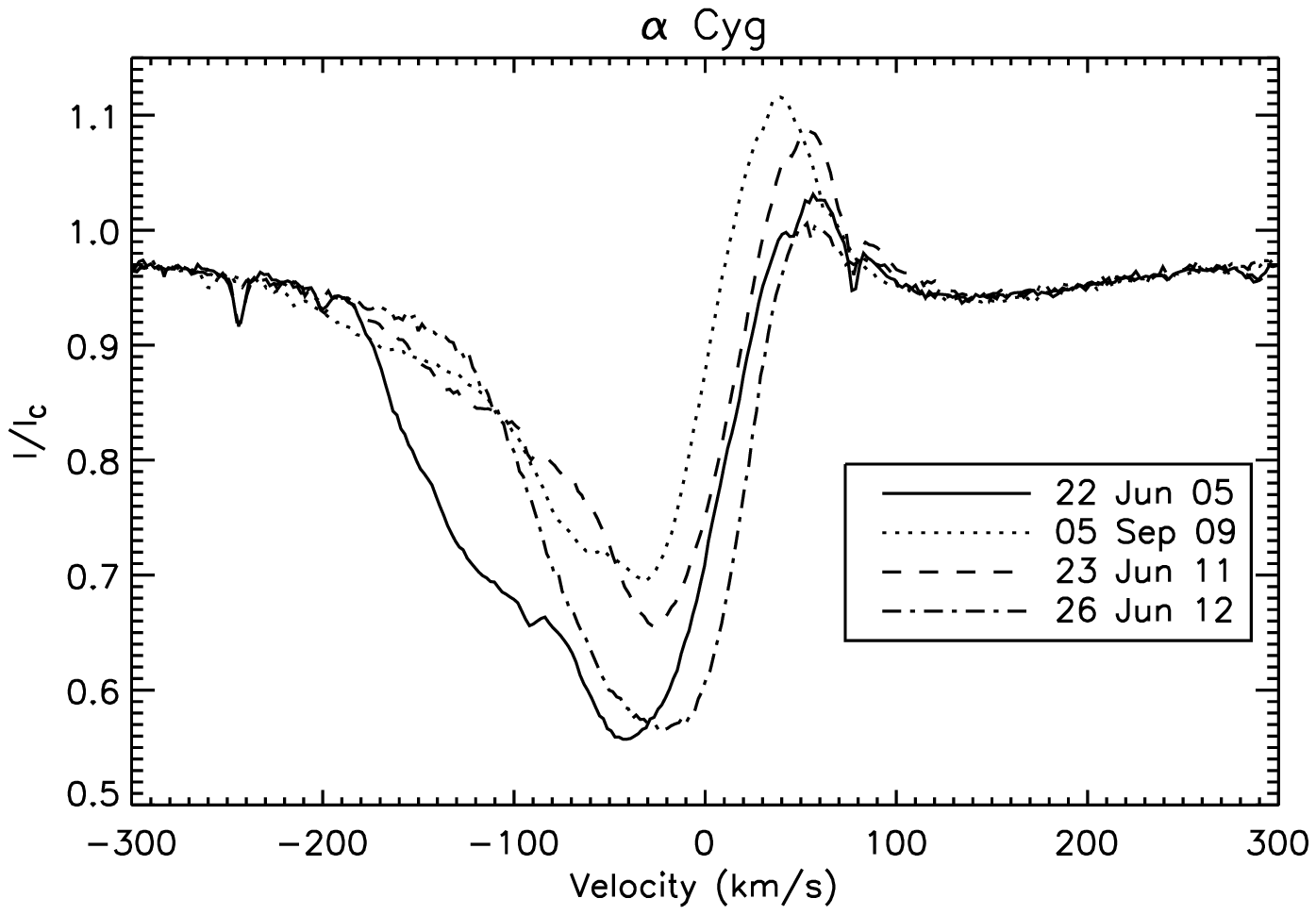} &
\includegraphics[width=2.25in]{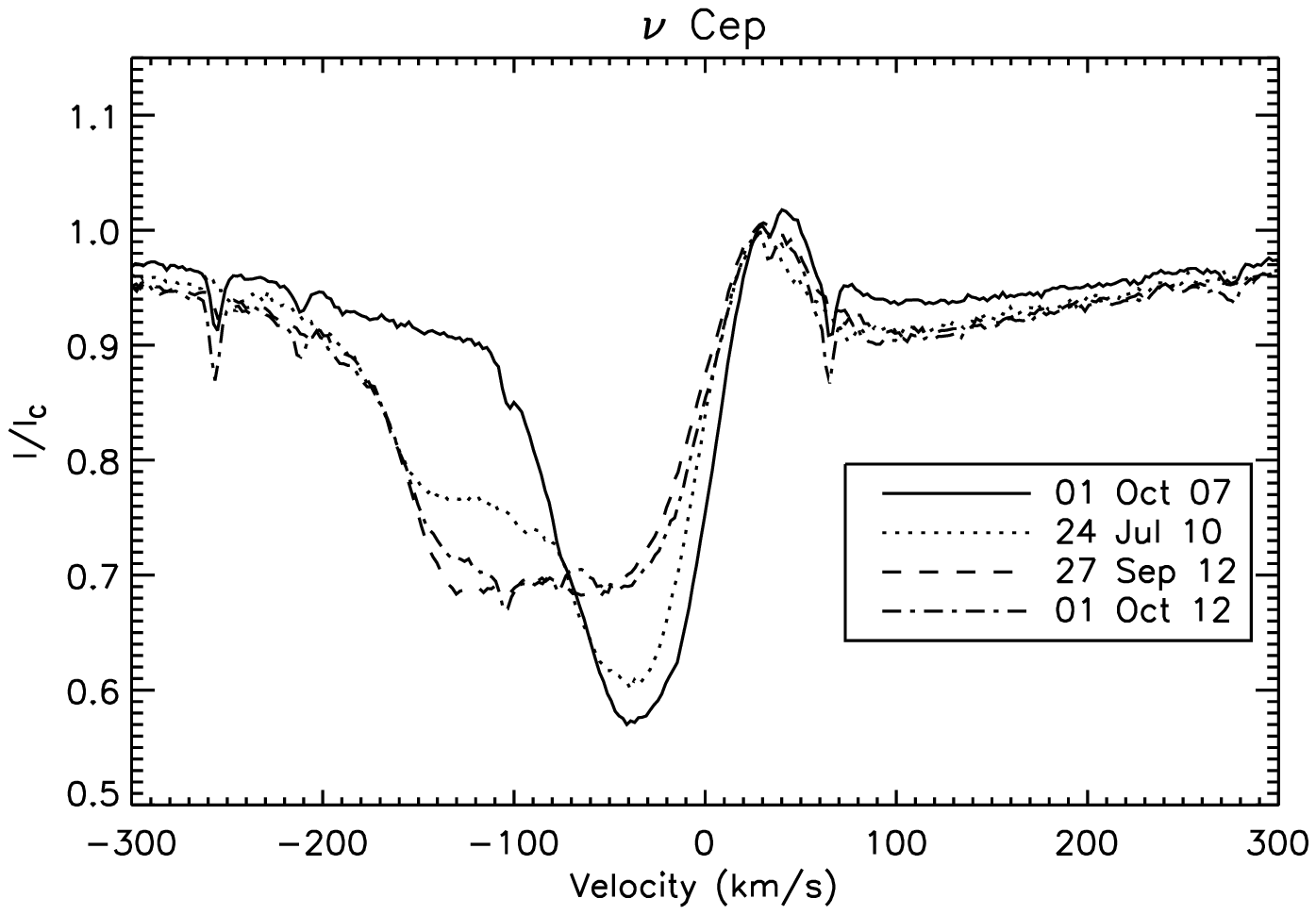} \\
\end{tabular}
\caption[H$\alpha$ profiles]{H$\alpha$ profiles of the 6 program stars. Different nights are plotted in different line-styles, with the observation dates given in the legends. In the case of $\beta$ Ori, due to the large number of observations, only a representative sample of line profiles is shown. $\alpha$ Cyg and $\nu$ Cep both show strong absorption features extending out to $-200$ \kms. The narrow absorption features evident in some spectra are due to telluric lines.}
\label{Halpha_profs}
\end{figure*}

\begin{figure*}
\begin{tabular}{cc}
\includegraphics[width=3in]{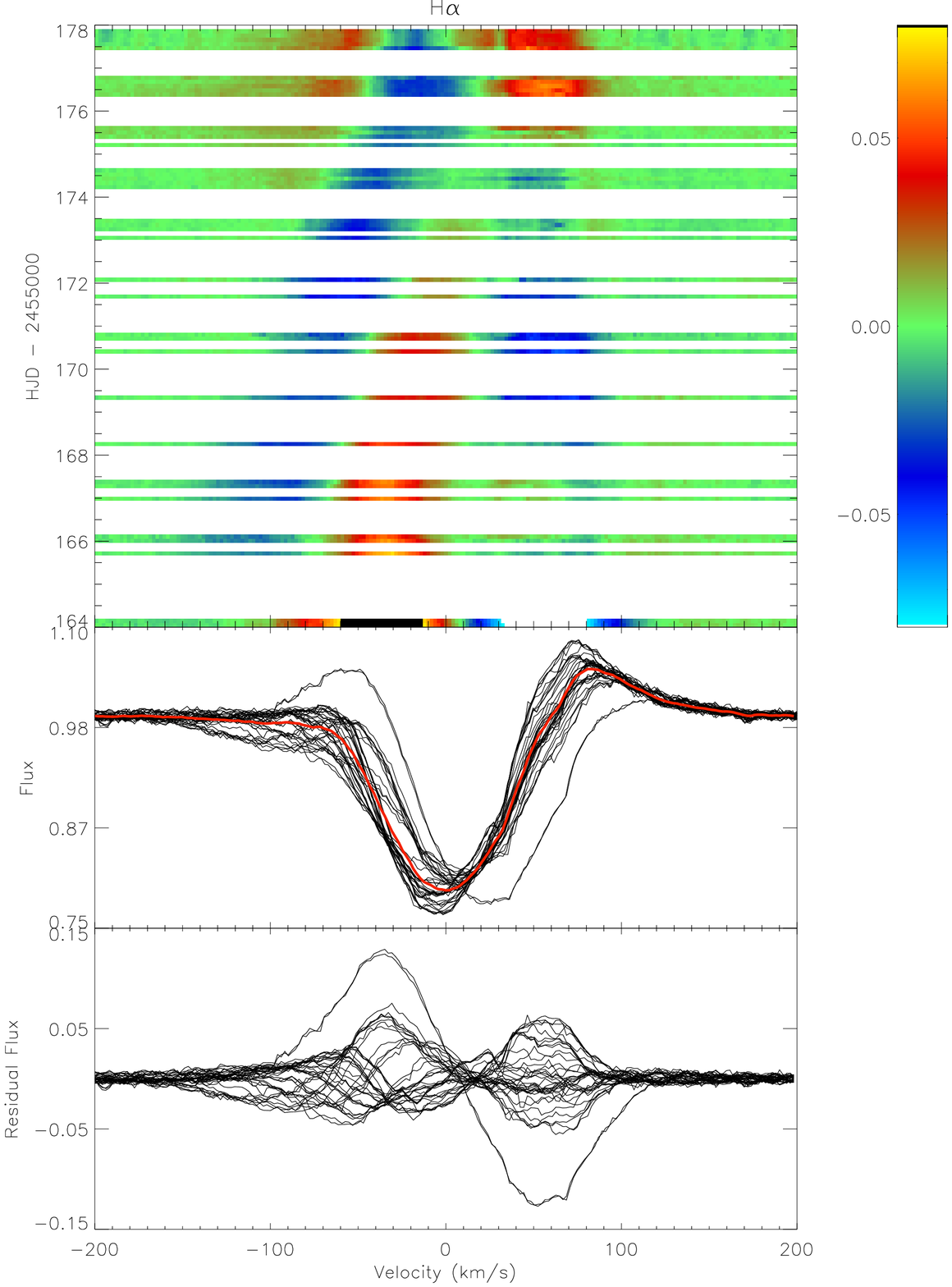}  &
\includegraphics[width=3in]{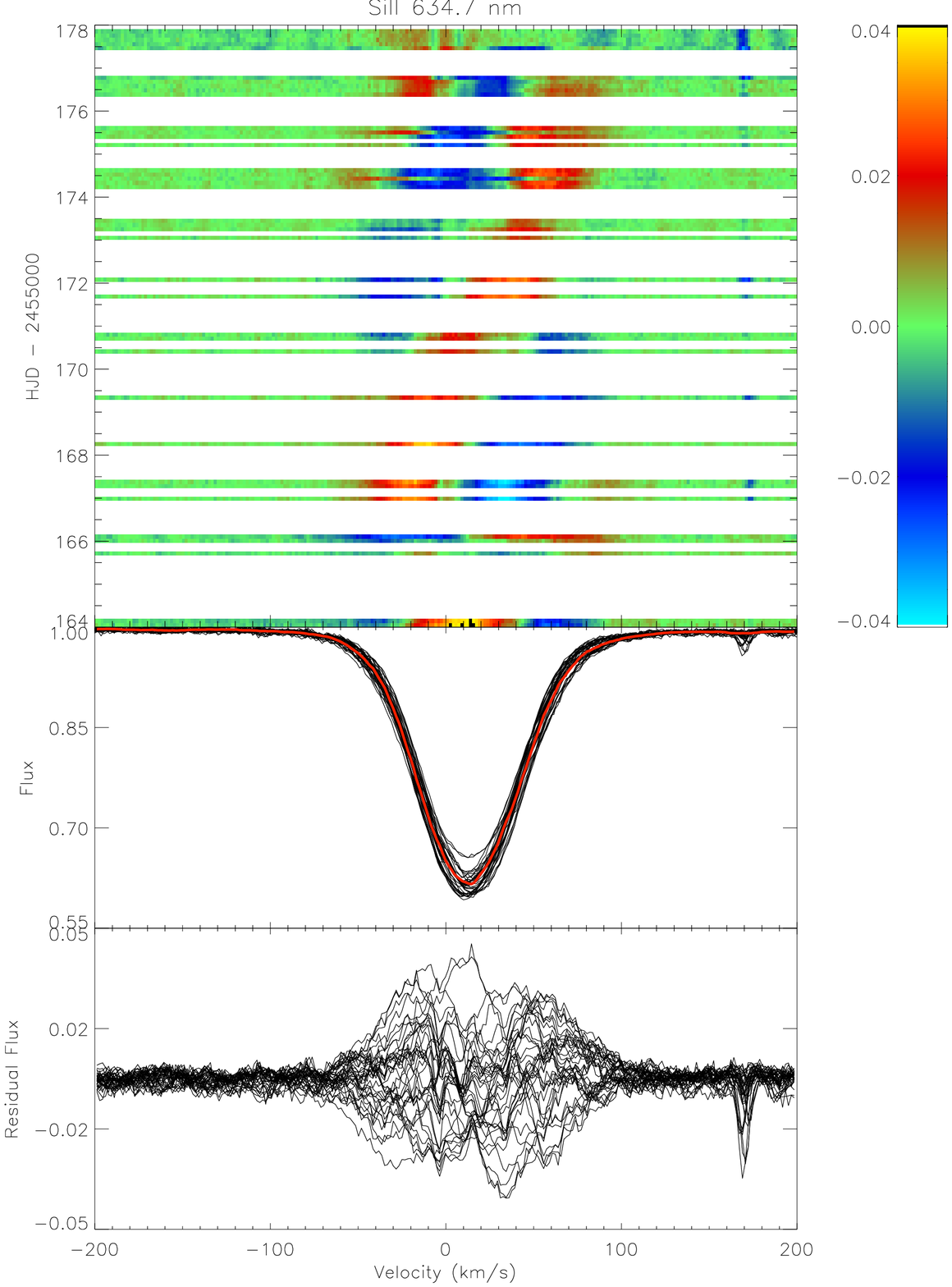}  \\
\end{tabular}
\caption[H$\alpha$ dynamic spectrum of $\beta$ Ori]{\textit{Top}: H$\alpha$ (left) and Si {\sc ii} 634.7 nm (right) dynamic spectra of $\beta$ Ori during the most densely time-sampled period, computed via the difference between individual spectra and the mean line profile. \textit{Middle}: one-dimensional intensity spectra. The red profile is the mean line-profile. \textit{Bottom}: residual intensity relative to the mean line profile. Note that the weak HVA, which appears on day 166 with a central velocity around $-100$ \kms, is preceded by an inverse P Cygni profile.} 
\label{balm_dyn}
\end{figure*}

With the exception of $\beta$ Ori a detailed time-series analysis of the targets is precluded by the paucity of data, however an attempt can still be made to characterize the wind activity during the observations. This is explored through the \halp~profiles shown in Fig. \ref{Halpha_profs}. Given the greater temporal sampling of $\beta$ Ori, only a representative selection of $\beta$ Ori's \halp~line profiles are shown; for the remaining stars all observations are overplotted. 

The \halp~profile of $\beta$ Ori is almost entirely filled with highly variable emission, demonstrating both P Cygni and inverse P Cygni profiles over the course of monitoring, as reported in previous investigations (e.g. \citealt{k1996b}). The star appears to have been relatively quiescent in comparison to some previous observing campaigns, showing spectral variability consistent with that seen in AST and BESO spectra \citep{mora2012a, shultz2012b, k2012}. In particular it does not show any spectacular HVAs. However, on Dec 1 2009 a weak absorption feature appeared around --150 \kms. The dynamic spectrum in Fig. \ref{balm_dyn}, which shows residual intensity compared to a mean line profile as a function of heliocentric Julian date (HJD) over the most densely time-sampled period, demonstrates that the feature appears quite suddenly following an inverse P Cygni profile, before migrating red-wards on a timescale consistent with the evolution of the HVAs previously observed for this star (around a month, \citealt{k1996b}). The dynamic spectrum of the Si {\sc ii} 634.7 nm line is shown in the second panel of Fig. \ref{balm_dyn}. Similarities between the two lines suggest that \halp~is also affected by non-radial pulsations.

While $\beta$ Ori does not show strong HVA activity in these observations, an overlapping ESO campaign captured the onset and subsequent evolution of a full HVA event approximately 20 days following the cessation of spectropolarimetric monitoring \citep{k2012}. Concurrent spectro-interferometry of the Br$\gamma$ line shows a differential phase signal that is at its strongest approximately 25 days before the onset of the 2012 HVA, corresponding to the final spectropolarimetric observations. \citeauthor{k2012} suggest that this may be due to an extended circumstellar structure rotating from next to the star (the differential phase signal) to in front of the star (the HVA). This would be consistent with a rotational period of $\sim$100 days.

The \halp~profile of 4 Lac is in full absorption, although there is some apparent variability around the line core.

The \halp~profile of $\eta$ Leo is in full absorption and shows virtually no variability, aside from a small amount in the blue-shifted half of the line. This is consistent with the pattern noted by previous investigations \citep{rich2011}. 

HR 1040 shows a double-peaked profile, consistent with previous observations \citep{verd1999, rich2011}. 

In the case of $\alpha$ Cyg, the first of the four observations shows a strong HVA event, while during other nights the star exhibits P Cygni profiles. HVAs are particularly rare for $\alpha$ Cyg \citep{rich2011}, indicating that this observation represents a fortunate opportunity to investigate the magnetic hypothesis for the origin of these events. 

In the case of $\nu$ Cep, we see an unremarkable P Cygni profile in the first observation, but a strong blue-shifted absorption feature is apparent in subsequent observations, with the most recent showing the strongest redistribution of flux absorption towards high velocities. Over multiple years of monitoring $\nu$ Cep has never been known to exhibit HVAs with an absorption depth comparable to those of $\beta$ Ori, however it does exhibit a highly variable blue-shifted absorption component \citep{rich2011}. While a detailed analysis of this star's wind variability based upon multi-year spectral monitoring has yet to be published, preliminary comparison of these observations to unpublished spectral monitoring suggests that the absorption feature is substantially stronger than average (Morrison, private communication). We therefore tentatively identify this feature as an HVA.

\section{Magnetic Analysis}

Magnetic fields are measured via the longitudinal Zeeman effect, as diagnosed through circular polarization induced in magnetically sensitive spectral lines. As a first step in the magnetic analysis, Least-Squares Deconvolution (LSD, \citealt{d1997, koch2010}) was performed on the individual spectra. LSD combines information from multiple lines to produce a mean line profile with a much higher SNR than any individual spectral line. To create this LSD profile, LSD deconvolves a `line mask' from the spectra: the line mask is a list of $\delta$ functions with locations and amplitudes corresponding to the wavelengths and continuum-normalized depths of spectral lines. Whenever possible a measured value for the Land\'e factor $g$ (a measure of the magnetic sensitivity of the line) is used; otherwise, this is calculated quantum mechanically assuming spin-orbit coupling. While the LSD Stokes $I$ profile is often an imperfect match to individual lines, the increase in the SNR of the Stokes $V$ profile has been repeatedly shown to enable the detection of magnetic fields that would otherwise have been far too weak to rise above the photon noise (e.g., \citealt{wade2000}). \cite{koch2010} provide a detailed discussion of the numerical implementation of LSD, together with an exploration of the inherent limitations arising due to the approximations necessary to the technique.

\begin{table*}
\centering
\caption[Longitudinal field measurements (coadded LSD profiles)]{Mean longitudinal field measurements from co-added spectra. The date corresponds to the average HJD of the individual spectra. The number of spectra used to create the profile is indicated by NS. $P_{\rm det}$ is the detection probability. Those observations corresponding to HVA events are indicated; the brackets in the case of $\beta$ Ori indicate the weak HVA-like behaviour discussed in the text. Bracketed sub-exposure times indicate those for Narval as supposed to ESPaDOnS observations. Magnetic measurements are described in more detail in Section 4.}

\begin{tabular}{|l|lcrrrlr|rr|rr|}
\hline
\hline
HJD & Calendar & HVA & subexp.& NS & LSD & $P_{\rm det}$ &\bz & \bz$/\sigma_B$ & \nz & \nz$/\sigma_N$ \\
--2450000 & Date & & time (s) &  & SNR  &   & (G) &   & (G) \\
\hline
\multicolumn{11}{c}{$\beta$ Orionis (87 lines)} \\
5100--5110 	&	 25 Sep--05 Oct 09 	&	 N 	&	 2 (5) 	&	3	&	8456	&	0.1549	&	 -10.3 $\pm$ 7.2 	&	-1.45	&	 7.8 $\pm$ 7.2 	&	1.08	\\
5115--5125 	&	 10--20 Oct 09 	&	 N 	&	 2 (5) 	&	4	&	10136	&	0.2544	&	 8.9 $\pm$ 6.1 	&	1.45	&	 3.9 $\pm$ 6.1 	&	0.63	\\
5125--5135 	&	 20--30 Oct 09 	&	 N 	&	 2 (5) 	&	6	&	10925	&	0.9829	&	 -4.8 $\pm$ 5.6 	&	-0.85	&	 -9.7 $\pm$ 5.6 	&	-1.72	\\
5167--5177 	&	 01--11 Dec 09 	&	 (Y) 	&	 2 (5) 	&	32	&	31451	&	0.8734	&	 2.8 $\pm$ 2.0 	&	1.39	&	 1.7 $\pm$ 2.0 	&	0.85	\\
5180--5190 	&	 14--24 Dec 09 	&	 N 	&	 2 (5) 	&	5	&	10023	&	0.0687	&	 1.7 $\pm$ 6.1 	&	0.27	&	 7.5 $\pm$ 6.1 	&	1.22	\\
5200--5210 	&	 03--13 Jan 10 	&	 N 	&	 2 (5) 	&	6	&	13214	&	0.3150	&	 2.8 $\pm$ 4.6 	&	0.61	&	 4.5 $\pm$ 4.6 	&	0.97	\\
5212--5222 	&	 15--25 Jan 10 	&	 N 	&	 2 (5) 	&	5	&	11704	&	0.9139	&	 8.1 $\pm$ 5.2 	&	1.55	&	 4.3 $\pm$ 5.2 	&	0.83	\\
5224--5234 	&	 27 Jan--06 Feb 10 	&	 N 	&	 2 (5) 	&	5	&	7383	&	0.2762	&	 10.3 $\pm$ 8.1 	&	1.27	&	 9.0 $\pm$ 8.1 	&	1.12	\\
5240--5250 	&	 12--22 Feb 10 	&	 (Y) 	&	 2 (5) 	&	2	&	5535	&	0.2394	&	 -4.1 $\pm$ 11.1 	&	-0.37	&	 7.4 $\pm$ 11.1 	&	0.67	\\
\hline
\multicolumn{11}{c}{4 Lacertae (76 lines)}  \\
6197.814	&	27 Sep 12	&	 N 	&	130	&	5	&	12915	&	0.0065	&	 2.0 $\pm$ 3.7 	&	0.53	&	 0.6 $\pm$ 3.7 	&	0.16	\\
6201.887	&	01 Oct 12	&	 N 	&	130	&	6	&	11339	&	0.0375	&	 0.5 $\pm$ 4.3 	&	0.11	&	 -7.0 $\pm$ 4.3 	&	-1.64	\\
\hline
\multicolumn{11}{c}{$\eta$ Leonis (196 lines)}  \\
4956--4959 	&	 04-07 May 09 	&	 N 	&	35	&	12	&	24036	&	0.9180	&	 -0.6 $\pm$ 1.8 	&	-0.35	&	 -0.4 $\pm$ 1.8 	&	-0.21	\\
5173.006	&	07 Dec 09	&	 N 	&	93	&	3	&	24783	&	0.1893	&	 -0.9 $\pm$ 1.8 	&	0.50	&	 -1.5 $\pm$ 1.8 	&	-0.81	\\
\hline
\multicolumn{11}{c}{HR 1040 (75 lines)}  \\
5405.010	&	27 Jul 10	&	 N 	&	182	&	3	&	7427	&	0.5967	&	 0.2 $\pm$ 4.5 	&	0.04	&	 -1.5 $\pm$ 4.5 	&	-0.34	\\
\hline
\multicolumn{11}{c}{$\alpha$ Cygni (466 lines)}  \\
3544.031	&	22 Jun 05	&	 Y 	&	15	&	1	&	8819	&	0.0560	&	 2.3 $\pm$ 3.4 	&	0.66	&	 2.8 $\pm$ 3.4 	&	0.83	\\
5079.813	&	05 Sep 09	&	 N 	&	11	&	3	&	23690	&	0.7711	&	 -0.4 $\pm$ 1.4 	&	-0.26	&	 -1.1 $\pm$ 1.4 	&	-0.74	\\
5736.096	&	23 Jun 11	&	 N 	&	2	&	16	&	24832	&	0.5764	&	 1.9 $\pm$ 1.3 	&	1.42	&	 -0.6 $\pm$ 1.3 	&	-0.48	\\
6105--6106 	&	 26--27 Jun 12 	&	 N 	&	5	&	80	&	76969	&	0.4813	&	 0.1 $\pm$ 0.5 	&	0.14	&	 0.4 $\pm$ 0.5 	&	0.91	\\
\hline
\multicolumn{11}{c}{$\nu$ Cephei (274 lines)}  \\
4374.900	&	01 Oct 07	&	 N 	&	75	&	1	&	7653	&	0.0233	&	 -2.9 $\pm$ 2.7 	&	-1.04	&	 -0.7 $\pm$ 2.7 	&	-0.24	\\
5402.129	&	24 Jul 10	&	 Y 	&	144	&	3	&	4789	&	0.0044	&	 -1.8 $\pm$ 4.3 	&	-0.42	&	 2.8 $\pm$ 4.3 	&	0.65	\\
6197.738	&	27 Sep 12	&	 Y 	&	100	&	6	&	23450	&	0.6783	&	 -2.9 $\pm$ 1.2 	&	-2.51	&	 0.5 $\pm$ 1.2 	&	0.91	\\
6201.846	&	01 Oct 12	&	 Y 	&	100	&	7	&	20392	&	0.8021	&	 -2.0 $\pm$ 1.3 	&	-1.53	&	 -0.4 $\pm$ 1.3 	&	-0.30	\\
\hline
\hline
\end{tabular}
\label{btab-mean}
\end{table*}

The line masks were adjusted for each star in order to ensure the LSD model was as faithful to the underlying spectra as possible. Emission lines, lines blended with Balmer or Paschen lines, interstellar lines, and telluric lines were not included in the mask as they introduce unacceptable distortion into the line profile, i.e. leaving only photospheric lines, which are not significantly affected by wind emission. In order to reduce noise arising from weak lines, lines with an observed central depth of less than 10\% of the continuum were also removed, since their contribution to the Stokes $I$ profile is primarily to make it noisier, while contributing insignificantly to the potential signal in Stokes $V$. The number of lines remaining in each mask are given in Table \ref{btab-mean}. The depths of the remaining spectral lines were empirically adjusted from their theoretical values so as to match the observed line depths, ensuring a better fit to Stokes $I$, and proper weighting of the contribution of each line to Stokes $V$.


In order to increase the SNR beyond that obtainable in a single observation, the LSD profiles of individual spectra were binned in time. Due to expected rotational modulation of any Stokes $V$ profile, binning cannot be performed over more than about 10\% of the rotational period without substantial distortion of any underlying Zeeman signature. While the rotational periods of these stars remain undetermined, upper and lower limits (see Table \ref{sad-tab-2}) indicate that they are likely to be on the order of 100 days for all stars (in the case of $\alpha$ Cyg, which may in fact be a `rapid' rotator seen close to pole-on \citep{ches2010}, the minimum period is $\sim$140 days). In no case but that of $\beta$ Ori are binned observations separated by more than three days. In the case of $\beta$ Ori a 10-day bin is adopted, which should correspond to approximately 10\% of the maximum rotational period. While it may be objected that this also corresponds to $\sim$1/3 of the \textit{minimum} rotational period, the preponderance of evidence suggests that $\beta$ Ori is seen at high inclination \citep{ches2010}, with a rotational period on the order of 100 days \citep{k2012}. 

It must be noted that $\beta$ Ori displays radial velocity variability on time scales of order 10 days, with an amplitude of $\sim$8 \kms. This is likely to be due to non-radial pulsations, as discussed by \cite{mora2012b}. Fig. \ref{balm_dyn} shows what may be an excited pulsational mode in the dynamic spectra for the Si {\sc ii} 634.7 nm line, but no sign of the higher velocity absorption components in the \halp~profile, lending confidence to the assumption that the photospheric metallic lines are essentially unaffected by the stellar wind (see also \citealt{k1996a, k1997}). This low-amplitude variability of the photospheric lines is small compared to that in other classes of stars in which magnetic fields are routinely detected, e.g. Ap/Bp stars, $\beta$ Cep stars, or Of?p stars (in which the variability arises due to photospheric abundance spots, pulsations, and magnetically confined winds, respectively), and so is not considered an impediment to detection of magnetic fields in this sample. Line profile distortion due to pulsation is approximately corrected for the purposes of the magnetic diagnosis by shifting each LSD profile to a common central velocity before adding them. Analysis of the pulsating magnetic star HD 96446 shows radial velocity correction improves the smoothness and increases the amplitude of Stokes $V$ profiles, thereby improving detectability, although some remaining line profile distortions are not accounted for by this method \citep{neiner2012}. Since the profiles are being combined across at least a full pulsation period, however, distortions in Stokes $V$ should cancel out. Radial velocity correction also reduces signatures in the null profile, although in this case there are no such signatures to begin with.

\begin{figure*}
\centering
\begin{tabular}{ccc}
\includegraphics[width=2.25in]{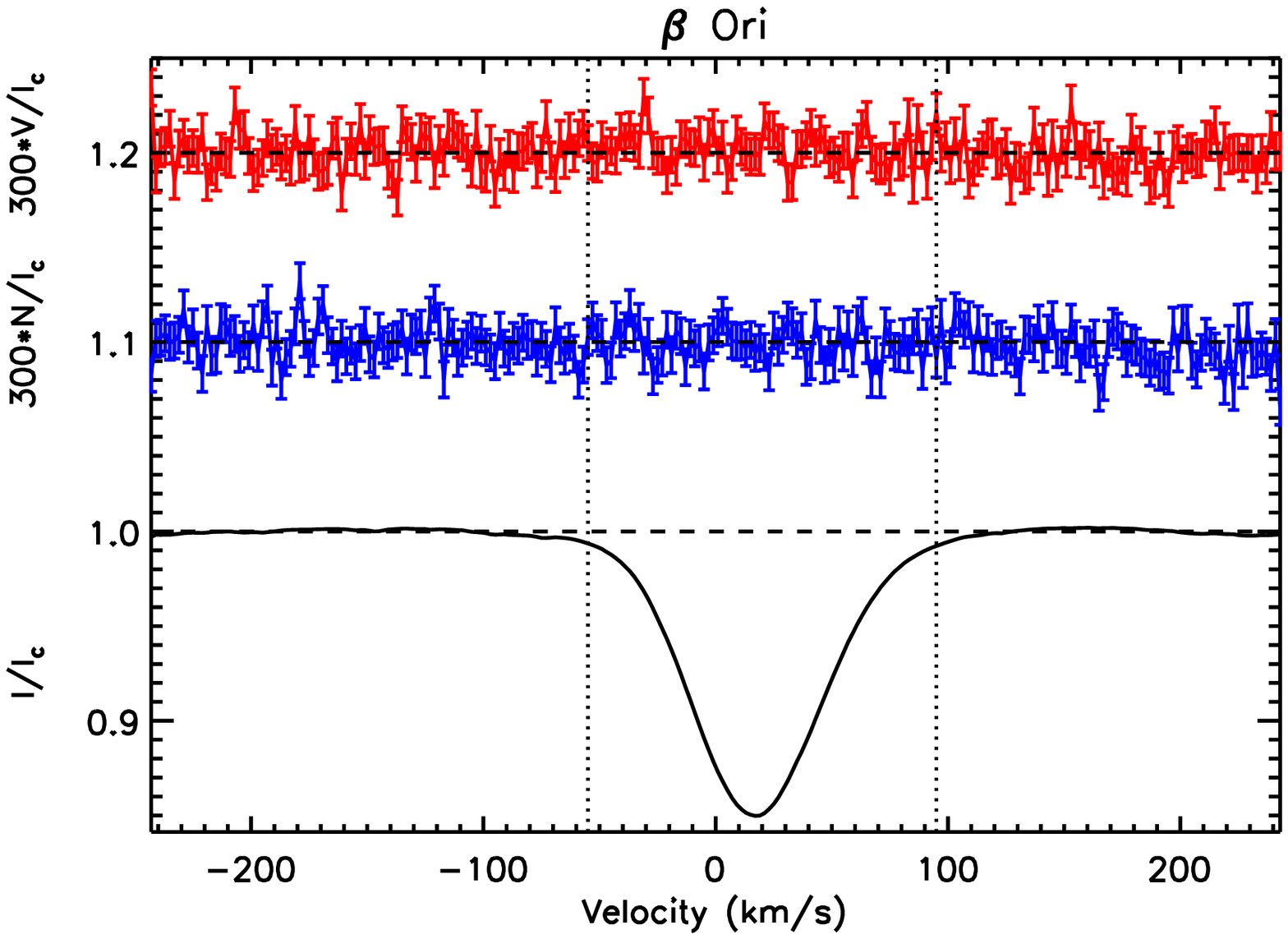} &
\includegraphics[width=2.25in]{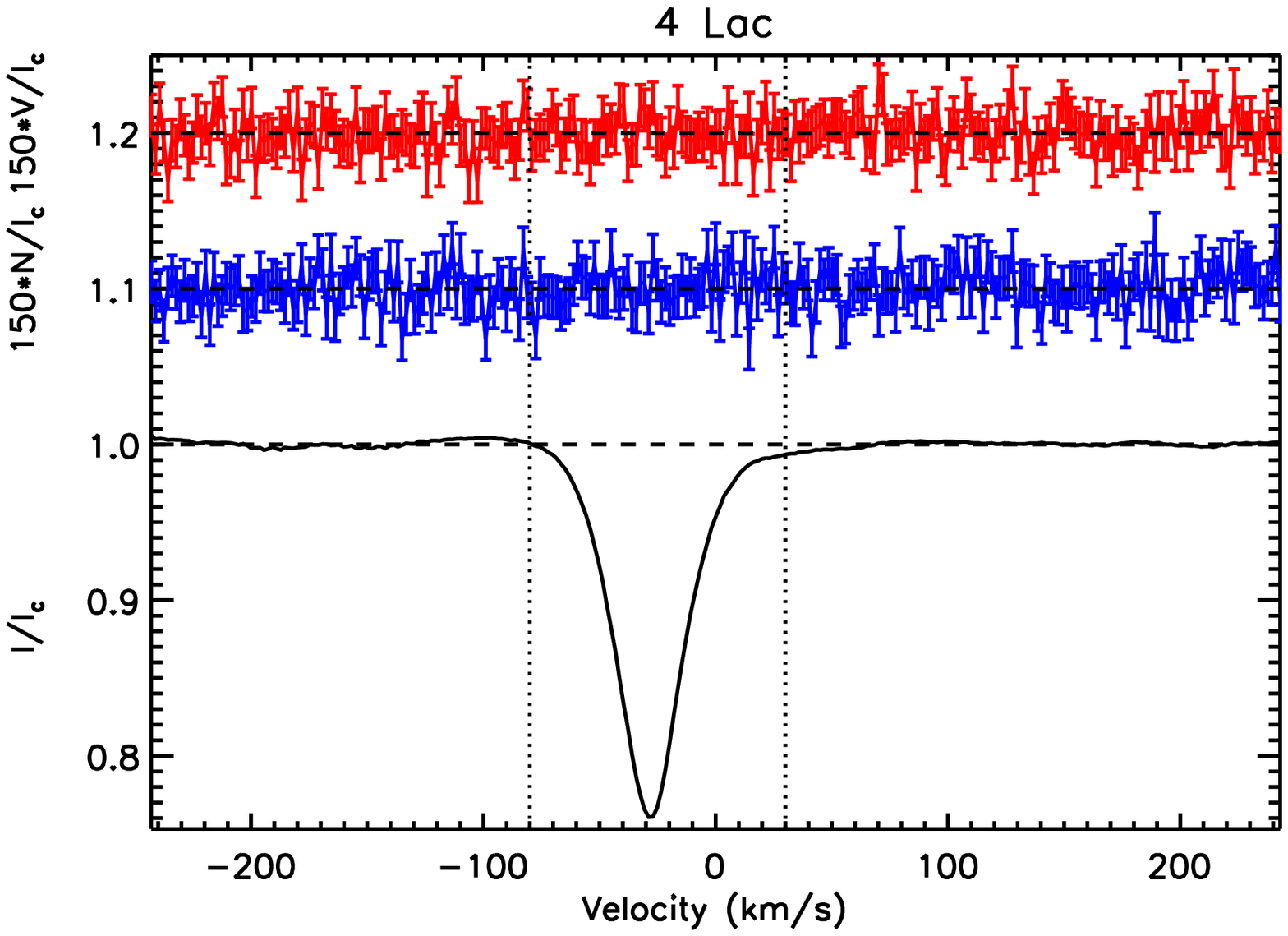} &
\includegraphics[width=2.25in]{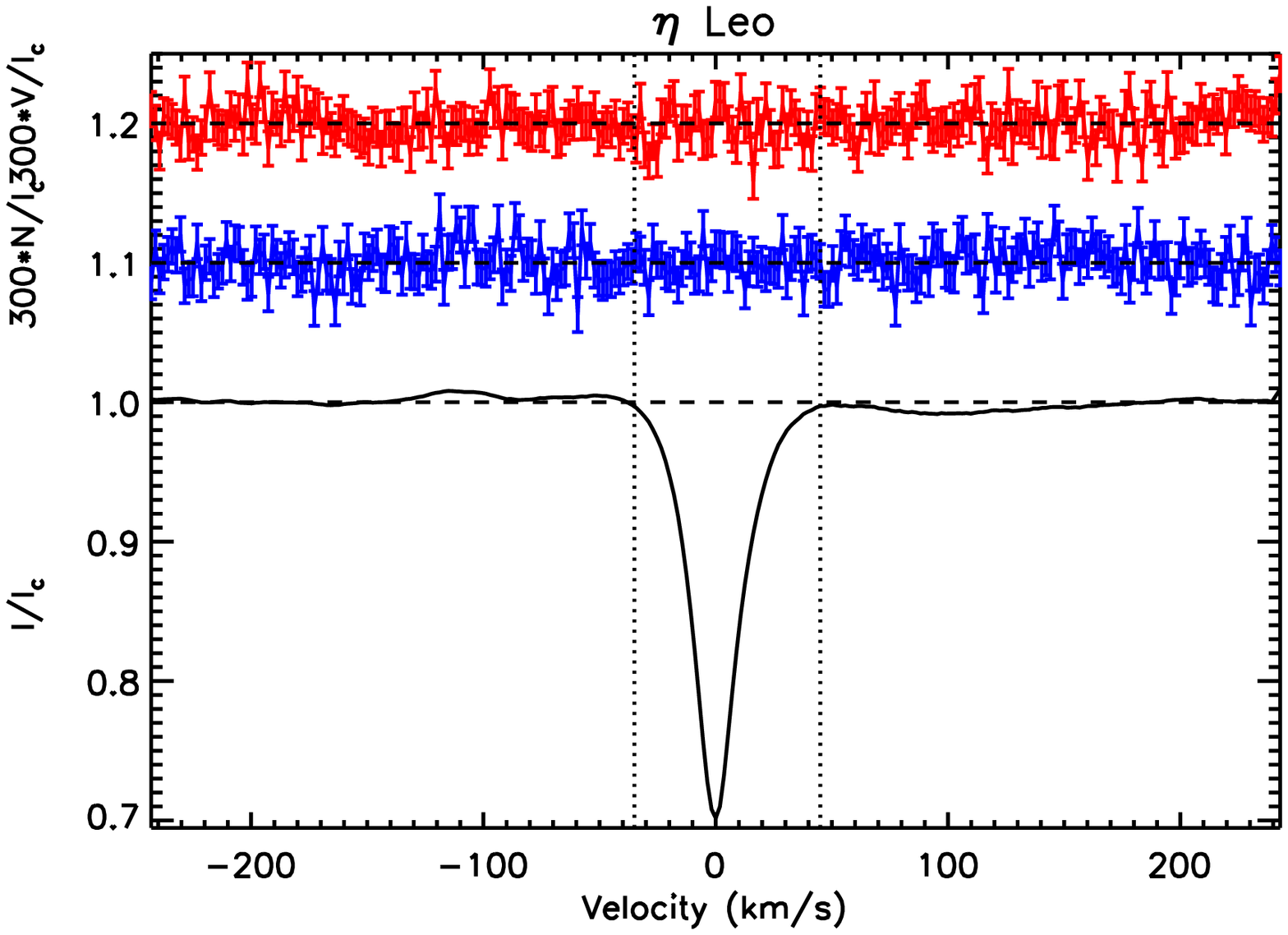} \\
\includegraphics[width=2.25in]{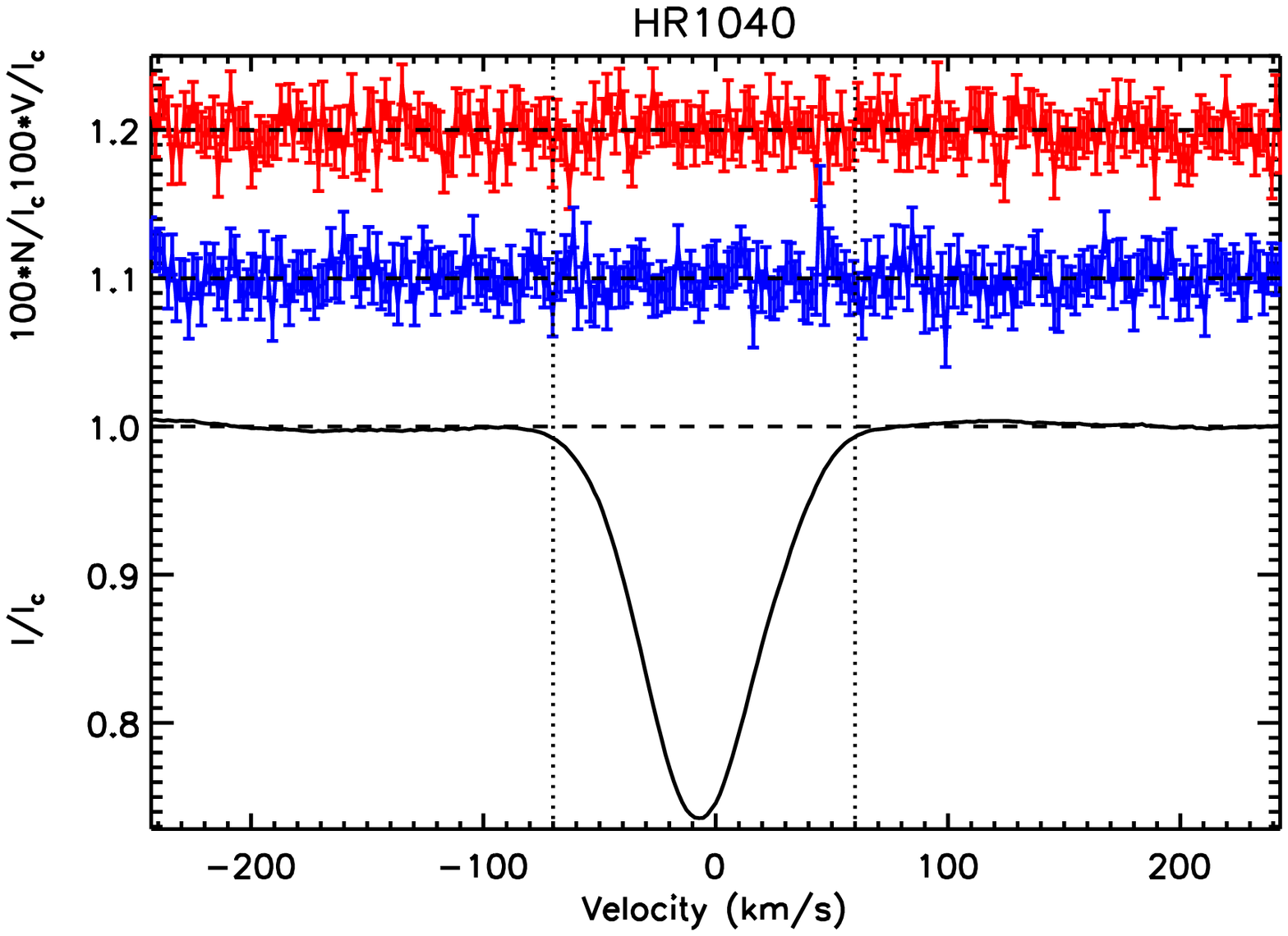} &
\includegraphics[width=2.25in]{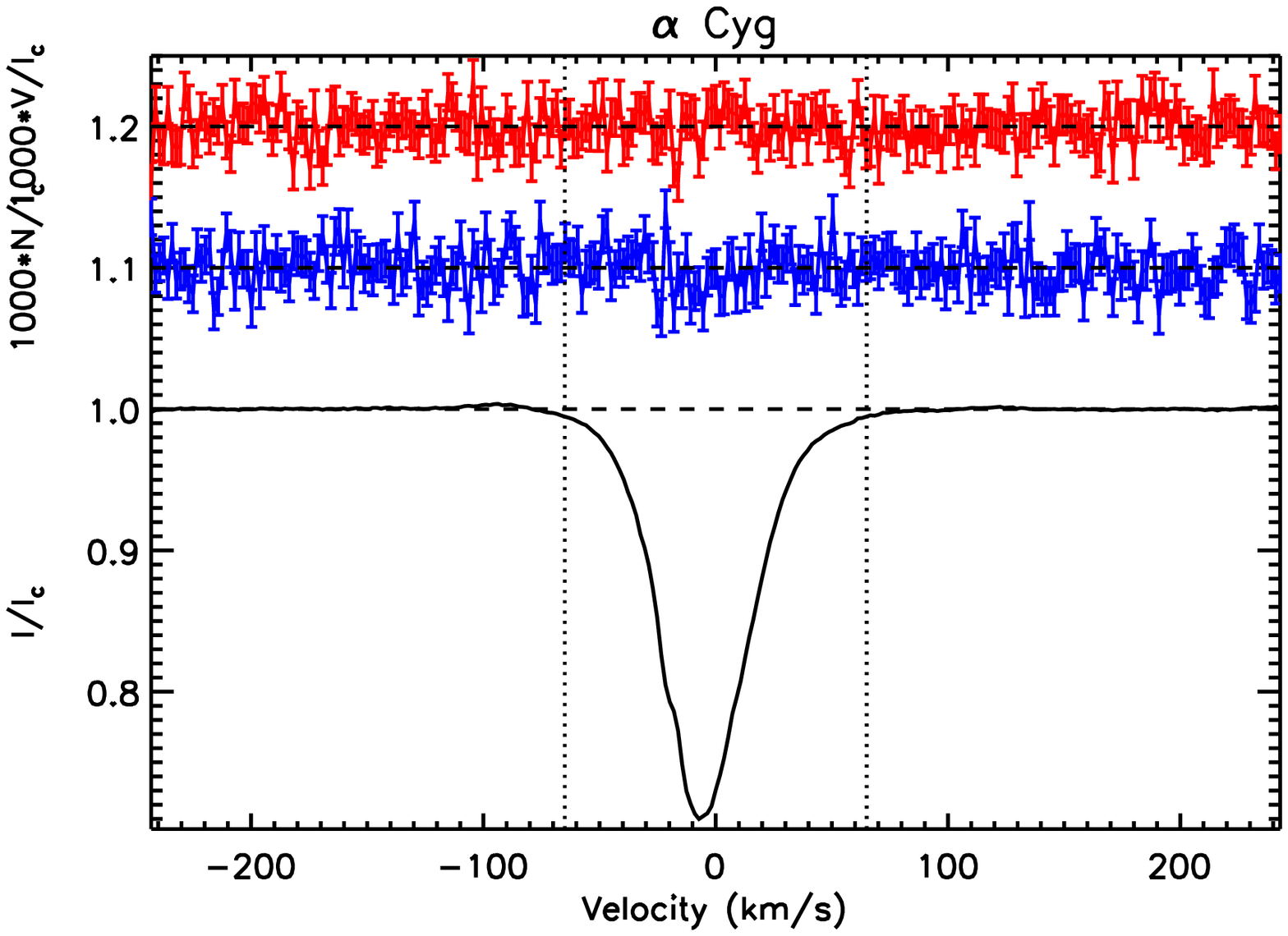} &
\includegraphics[width=2.25in]{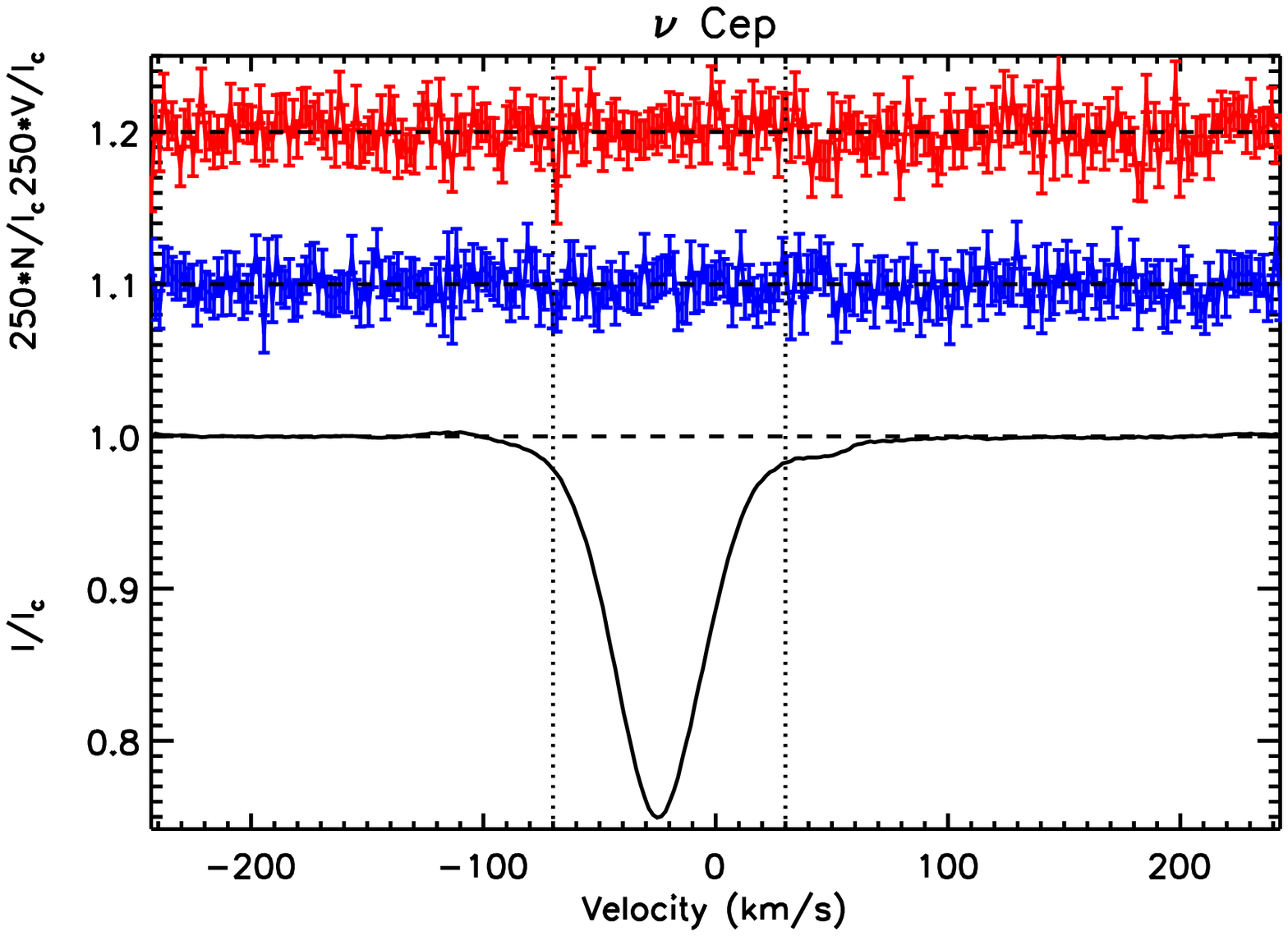} \\
\end{tabular}
\caption[LSD profiles]{LSD profiles for the six program stars showing Stokes $V$ (top, red), diagnostic $N$ (middle, blue) and Stokes $I$ (bottom, black). $V$ and $N$ profiles are amplified by a factor given in the axis titles. Vertical dotted lines represent the integration ranges.}
\label{lsd_coadd}
\end{figure*}

The co-added LSD profiles with the highest SNRs are shown in Fig. \ref{lsd_coadd}. The number of spectra combined to compute each LSD profile, the dates of observation, and the resulting SNRs are tabulated in Table \ref{btab-mean}. 

Two magnetic diagnostics were applied to each LSD profile: a statistical test of the significance of the polarized signal within the line profile, and measurement of the mean longitudinal magnetic field. The former searches for significant departure of the Stokes $V$ profile within the spectral line from the noise spectrum determined from both the $V$ profile outside the spectral line and the $N$ profile \citep{dsr1992}. A magnetic signature is considered `definite' if the formal detection probability $P_{{\rm det}}$ is greater than 99.999\% inside the $V$ line profile and negligible elsewhere; `marginal' if the detection probability is between 99.9\% and 99.999\%; and a non-detection otherwise. This test results in a null detection for all LSD profiles computed in this investigation (see Table \ref{btab-mean}). \textit{Hence we conclude that no magnetic fields are detected in any of our sample stars.}

The brightness-weighted, disk-averaged longitudinal magnetic field is computed from the first-order moment of the Stokes \textit{V} profile within the line (see e.g. \citealt{mat1989}): 

\begin{equation}\label{blong}
\langle B_z  \rangle  = -2.14\times 10^{11}\frac{\int \! vV(v)\mathrm{d}v}{\lambda g_{\rm{eff}}c\int \left[I_{\rm{c}}-I(v)\right] \mathrm{d}v}
\end{equation}

\noindent where \textit{v} is the velocity (in \kms) within the profile measured relative to the centre of gravity, and $\lambda$ and $g_{{\rm eff}}$ are the reference values of the wavelength (in nm) and the effective Land\'e factor used in computing the LSD profiles. The longitudinal field $\langle N_z\rangle$ of the null profile was also computed. The 1-$\sigma$ uncertainties associated with $\langle  B_z\rangle$ and $\langle  N_z\rangle$ were determined by propagating the uncertainties of each LSD pixel through Eq. \ref{blong}. The integration ranges used for these measurements, and for the statistical test described above, are shown in Fig. \ref{lsd_coadd}. Error bars are a function of both the SNR and the integration range, where the latter depends on line broadening: thus, more rapidly rotating stars will have larger error bars at a given SNR.

The quantities \bz~and \nz~for the ESPaDOnS and Narval observations are tabulated in Table \ref{btab-mean}, together with the ratio of \bz~and \nz~to their error bars. All magnetic field measurements in all observations are beneath the 3$\sigma$ significance threshold, with error bars ranging from 0.5--11.5 G and a mean error bar of 4.3 G. In agreement with the statistical tests, no significant longitudinal magnetic field is detected in any of the stars.

The co-added LSD profiles for all stars were used as input for the Bayesian search algorithm developed by \cite{petit2012a}, in order to constrain the strength of dipole component of the stellar magnetic field. Modeling the magnetic field as a dipole and generating synthetic LSD profiles for varying dipolar field strength $B_{\rm d}$, obliquity of the magnetic axis $\beta$, and inclination of the rotational axis $i$, the approach yields probable upper limits on undetected underlying stellar dipolar magnetic fields. This method is able to establish robust constraints with a small number of observations \citep{petit2012a}. While the upper limits are obviously affected by the star's intrinsic line broadening and the SNR of individual measurements, the number of observations also plays an important role: with only a small number of observations, a significant probability tail persists at high $B_{\rm d}$, which diminishes as further observations are added.

The cumulative probability of the resulting probability density functions (PDFs), marginalized over $i, \beta$, and $B_{{\rm d}}$, are shown in Fig. \ref{bd_lim}, with the 95.4\% credibility region filled in red. Comparison of synthetic $N$ and Stokes $V$ profiles has shown this to represent the threshold above which definite detection of the field becomes likely \citep{petit2012a}. The value of $B_{\rm d}$ corresponding to 95.4\% credibility is given in Table \ref{eta_bd_lim}, along with the logarithmic odds ratios of $M_0$, the null hypothesis, to $M_1$, the dipolar magnetic field hypothesis. If $\log{(M_0/M_1)} \simeq 1$, as is the case for all stars, there is no evidence favouring a dipolar field over no magnetic field.

\section{Discussion}

\subsection{Magnetic Wind Confinement}

As demonstrated by \cite{ud2002}, the interaction between an outflowing line-driven stellar wind and a large-scale magnetic field can be characterized by the magnetic wind confinement parameter $\eta_*$, which is the ratio of magnetic to kinetic energy density at the stellar surface, given by 

\begin{equation}\label{etastar}
\eta_{*}= \frac{B_{\rm eq}^{2}R_{*}^{2}}{\dot{M}v_{\infty}} 
\end{equation}

\noindent where $R_*$ is the stellar radius, $\dot{M}$ is the mass loss rate, \vinf~is the terminal wind velocity, and $B_{\rm eq} = B_d/2$ is the equatorial magnetic field strength (all in cgs units), where the radial wind flow most directly opposes the horizontal orientation of the magnetic field. 

If $\eta_* > 1$, the wind will be magnetically confined: the outflowing wind plasma flows along closed field lines above the stellar surface to collect at the magnetic equator, where it is channeled into a wind shock of dense plasma radiatively cooled via X-ray emission (as has been proposed and observed amongst some magnetic early-type stars, e.g. $\theta^1$ Ori C, \citealt{bm1997, d2002}).

Measurements of \mdot~are available for most of the sample stars. Alternatively, theoretical mass-loss rates could be used. Both approaches suffer from a similar weakness, namely the application of a spherically-symmetric model to systems that are clearly not spherically symmetric \citep{mark2008}. In addition, there are large discrepancies between theoretical and empirical mass-loss rates for BA SG stars, and between determinations of \mdot~from different spectral regions. Whether these discrepancies are a consequence of wind geometry, of systematic errors in the theoretical calculations, or of systematic errors in the translation of spectrum to \mdot, is unclear.

In their study of magnetic wind confinement amongst magnetic early-type stars, \cite{petit2013} adopted theoretical mass-loss rates calculated with the empirically-calibrated recipe of \cite{vink1999, vink2000, vink2001}. As the recipe of \citeauthor{vink1999} is computed only for stars with $T_{\rm eff}\ge 12500$ K, application to BA SG stars requires extapolation. Although this formulation should be valid down to $\sim$7000 K, where molecular and dust effects become important (Vink, priv. communication), the predicted mass-loss rates for BA SG stars are 1--2 orders of magnitude higher than measured values (e.g. \citealt{BarlowCohen1977, mark2008a, ches2010}).

In recognition of the large discrepancy between observed and predicted mass-loss rates, we provide constraints using empirically inferred values of \mdot. However, it should be kept in mind that the empirical mass-loss rates are inferred using spherically-symmetric models applied to a variety of observational diagnostics: near-infrared photometry \citep{BarlowCohen1977}, \halp~profile modeling \citep{schil2008, mark2008}, and optical and near-infrared interferometry \citep{ches2010}. Therefore, to enable a consistent comparison across the sample, we also compute the theoretical mass loss rates of \cite{vink1999}, calculated using the parameters in Table \ref{sad-tab-2} and solar metallicities (a reasonable approximation for these stars, see e.g. \citealt{prz2006} and \citealt{schil2008}).

As all stars are below the second bistability jump \citep{vink1999}, wind terminal velocities were computed according to the relationship \vinf~=~$0.7v_{\rm esc}$, an empirical formula derived from the UV P Cygni troughs of a sample of OBA stars by \cite{lamers1995} and employed by \citeauthor{vink1999} If instead measured values of \vinf~are used, wind momentum changes by $\sim1/3$: however, as $\eta_*$ is sensitive to $B^2$, the resulting decrease in the magnetic field strength required to confine the wind is modest (a few G), and does not strongly affect the conclusions.

\begin{figure*}
\centering
\begin{tabular}{ccc}
\includegraphics[width=2.25in]{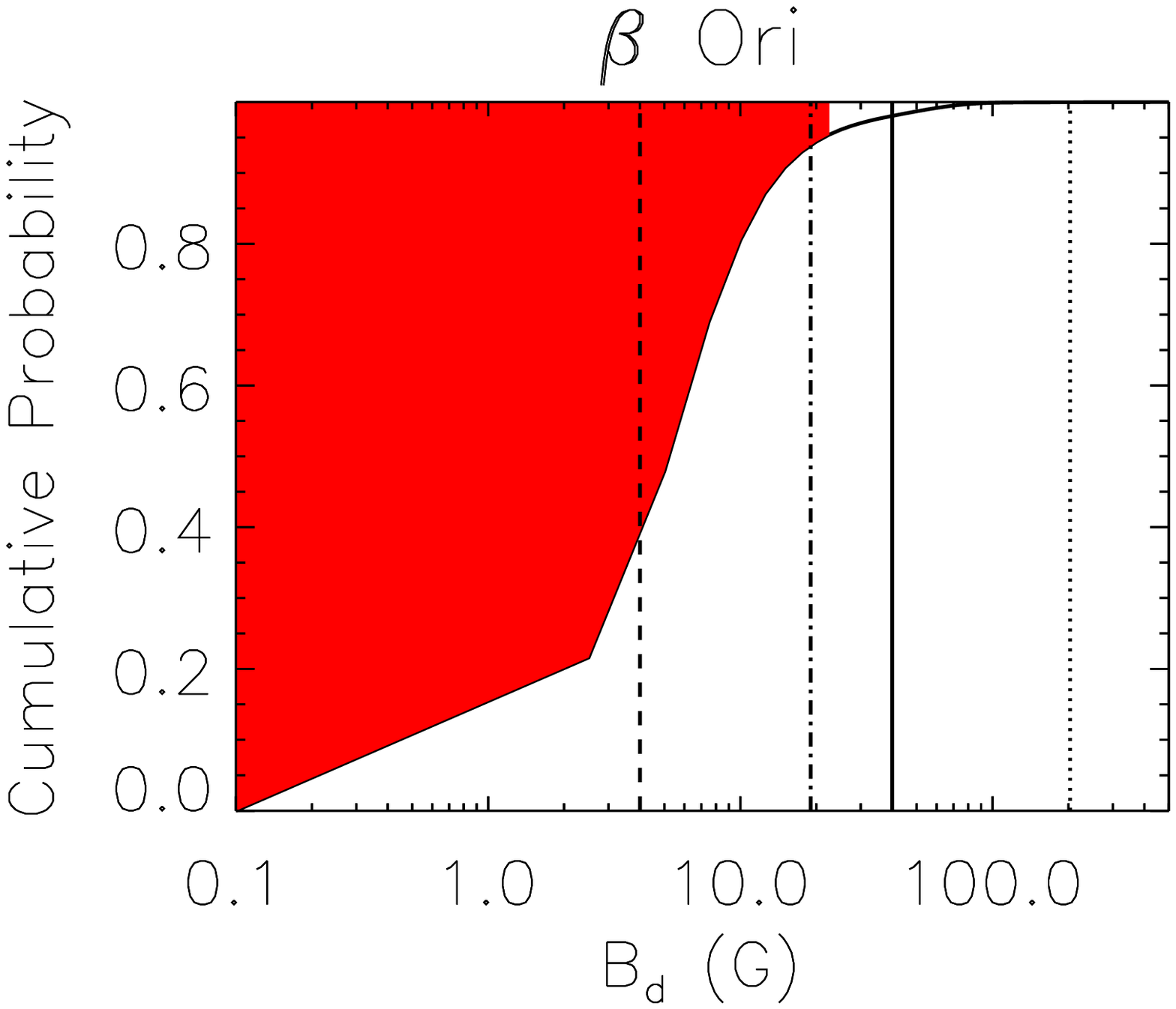} &
\includegraphics[width=2.25in]{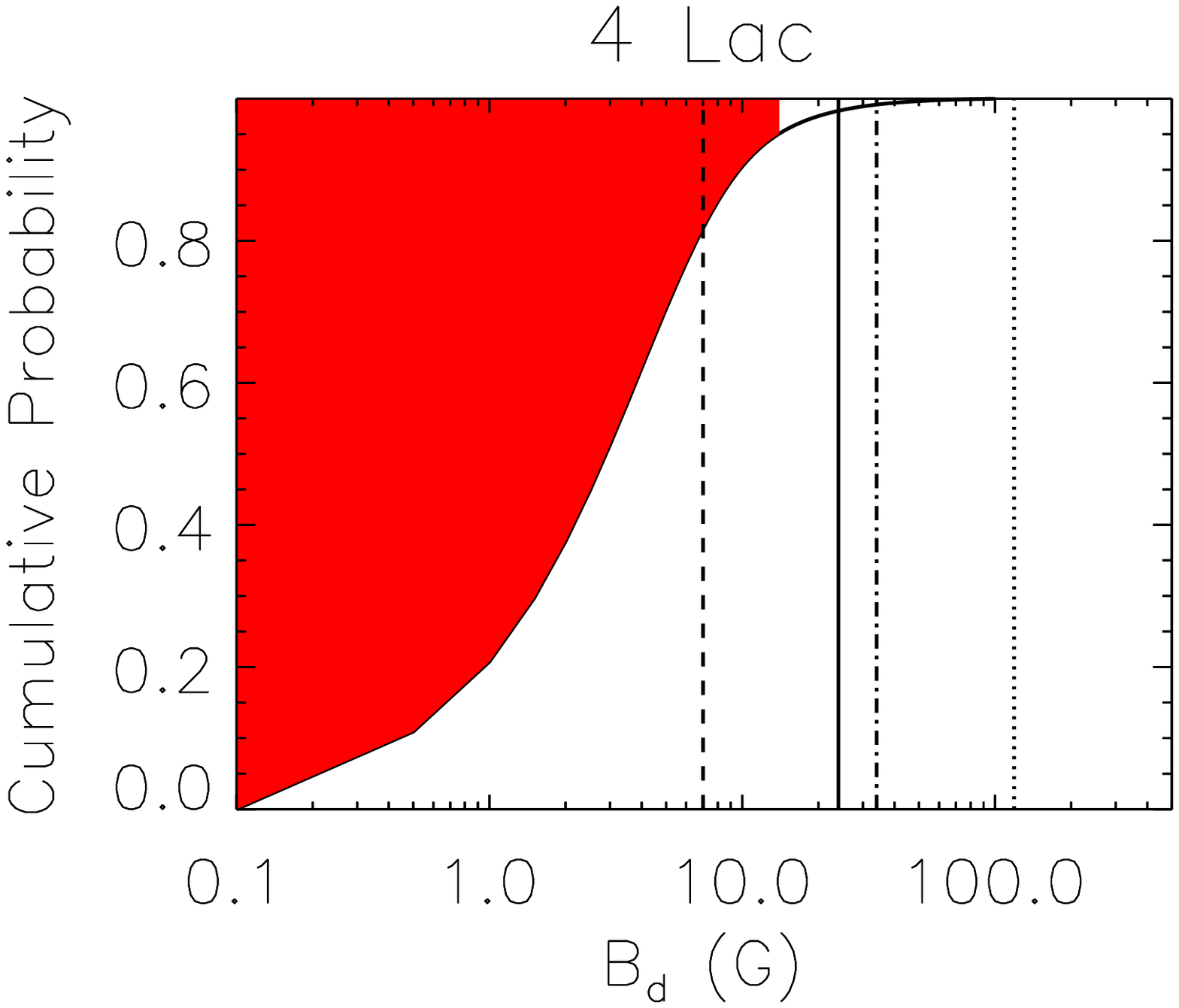} &
\includegraphics[width=2.25in]{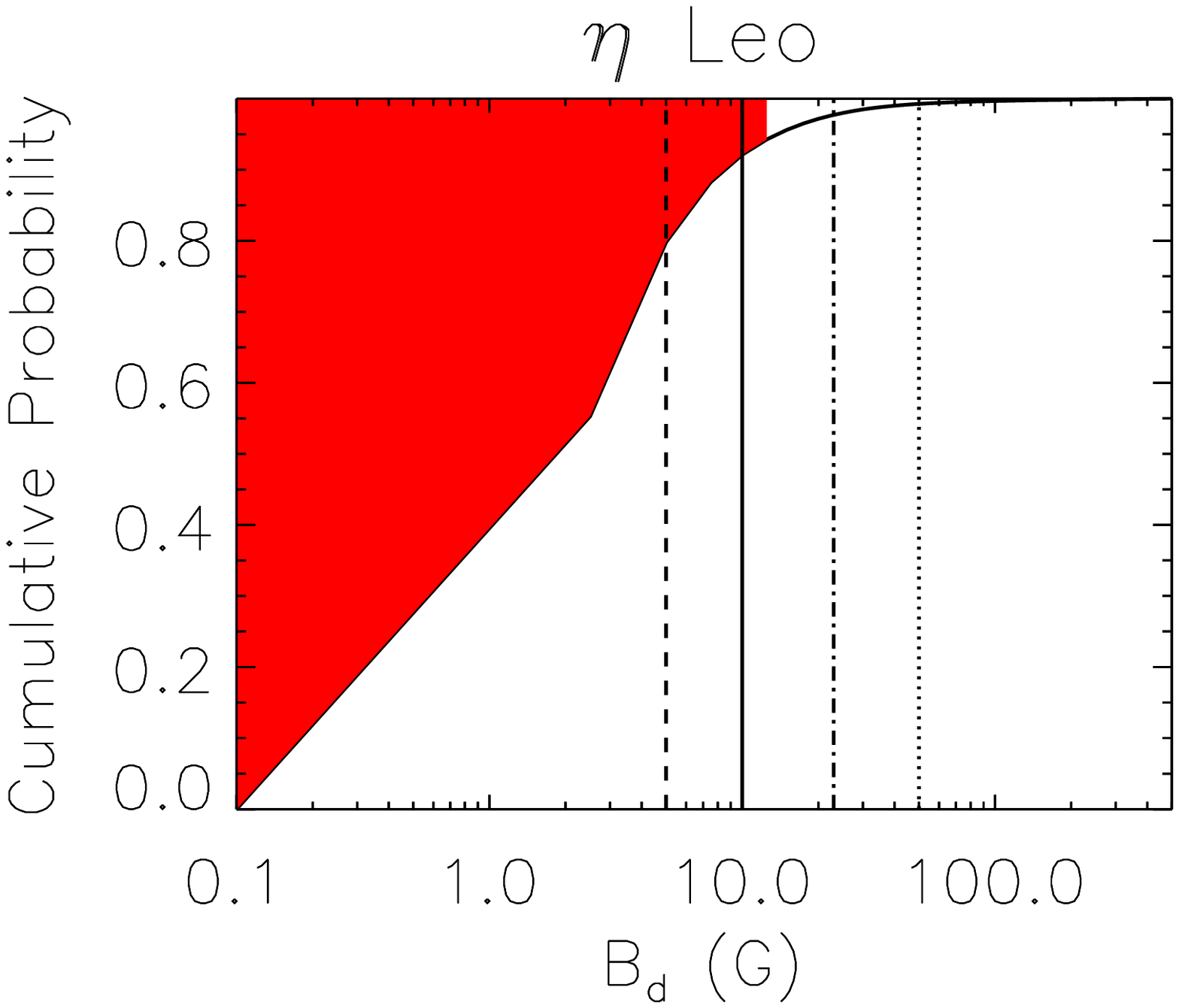} \\
\includegraphics[width=2.25in]{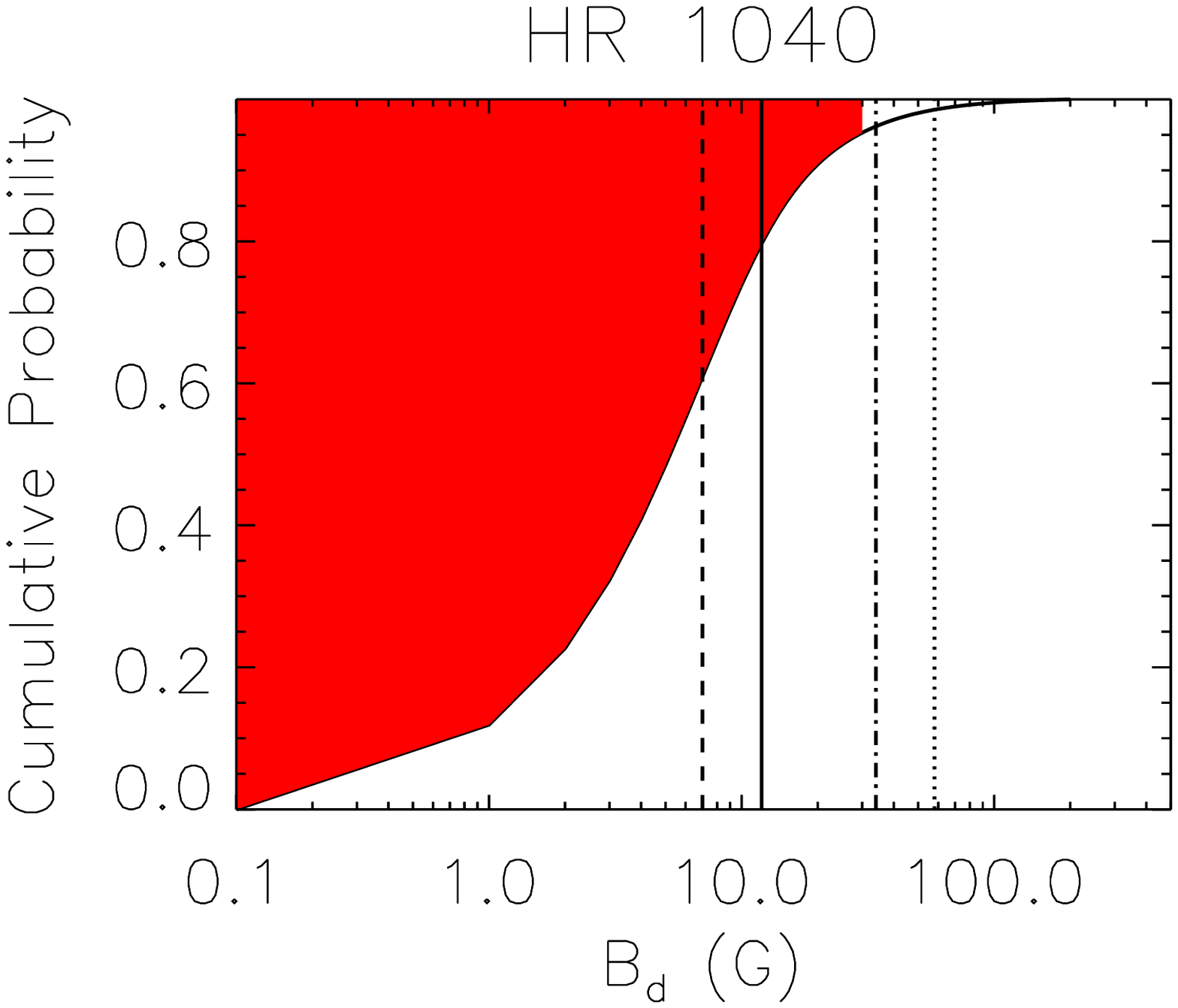} &
\includegraphics[width=2.25in]{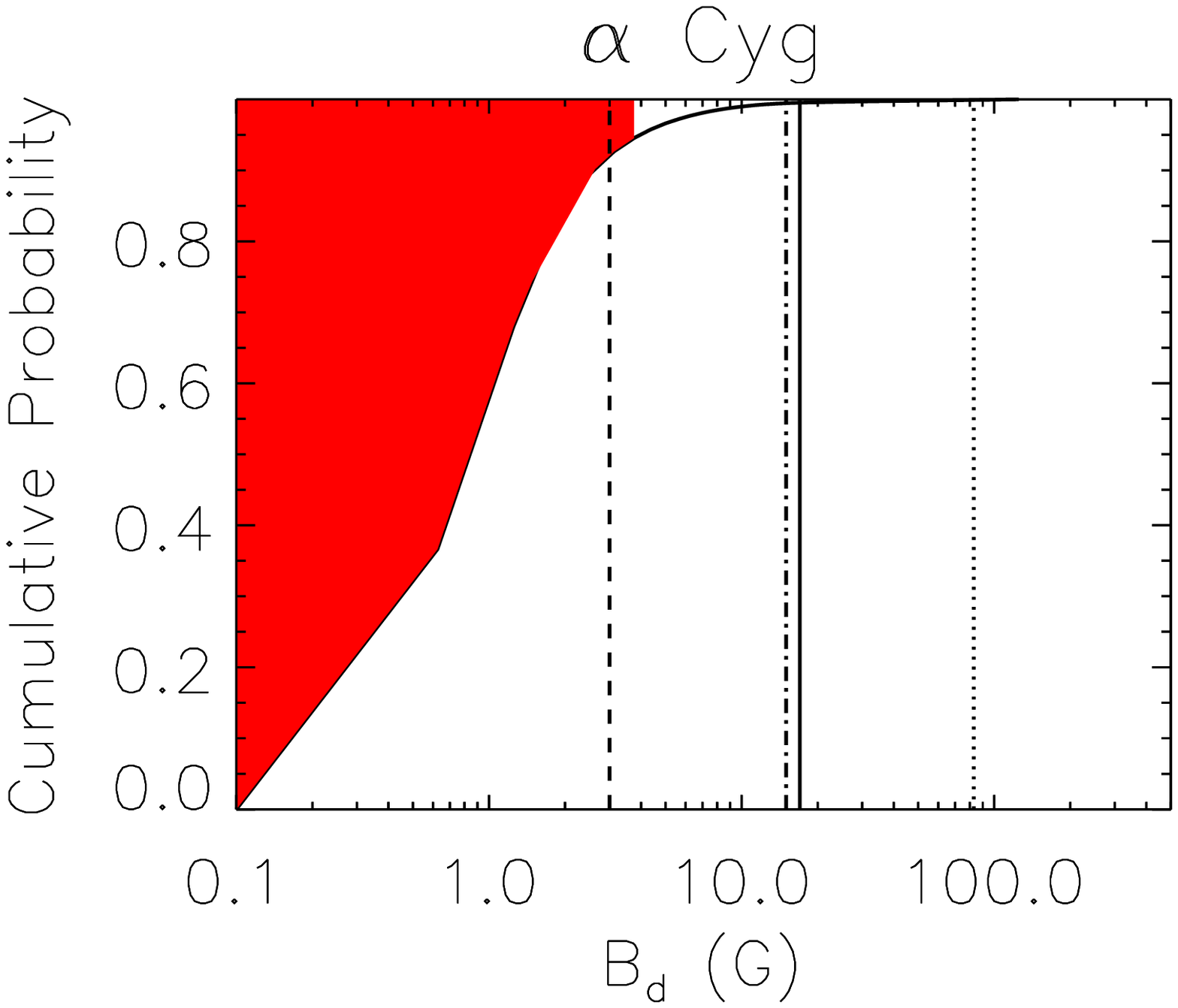} &
\includegraphics[width=2.25in]{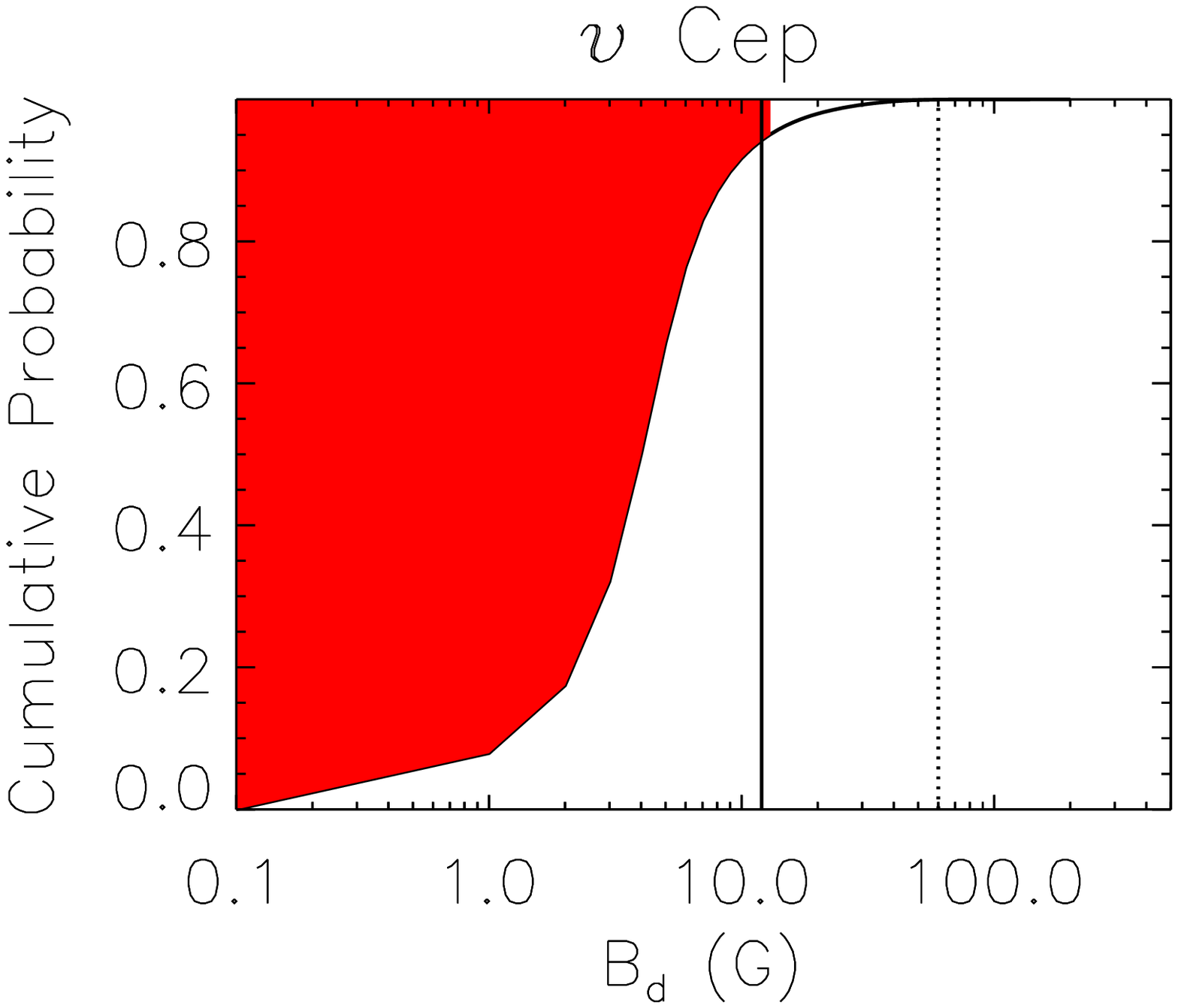} \\
\end{tabular}
\caption[Dipolar upper limits]{Cumulative probability density of the dipolar field established with Bayesian modeling of LSD Stokes $I$ and $V$ profiles. There is a 95.4\% probability of detecting fields with $B_d$ greater than that indicated by the filled (red) portions of distributions. Solid lines indicate the dipolar field strength required for $\eta_*=1$, dotted lines for $\eta_*=25$, using theoretical values for \mdot; dashed and dot-dashed lines indicate field strengths for the same values of $\eta_*$ using measured values of \mdot.}
\label{bd_lim}
\end{figure*}

The dipolar field strengths required for a confinement $\eta_*=1$, using theoretical mass-loss rates, are indicated by the solid lines in Fig. \ref{bd_lim}; dashed lines indicate the field strengths required according to empirical \mdot. For $\beta$ Ori, $\alpha$ Cyg, $\nu$ Cep, and 4 Lac, a magnetic dipole of sufficient polar strength to confine the wind in closed loops above the stellar surface is ruled out with a credibility greater than 95.4\% if theoretical mass-loss rates are used. The relatively weak winds of $\eta$ Leo and HR 1040 mean that constraints are not as good for these stars. With measured mass-loss rates, $\eta_* = 1$ cannot be ruled out for any of the stars.

The condition $\eta_*=1$ is simply the threshold for magnetic confinement: in order to reproduce the variability seen in these stars, a larger value might be required. Based upon \halp~visibilities, \cite{ches2010} suggest that the emission line formation regions of $\alpha$ Cyg and $\beta$ Ori are extended to $\sim 1.5-1.75 ~R_*$ from the stellar centre; differential phase signal variability additionally suggested significant perturbations in $\alpha$ Cyg's wind at 2--3 $R_*$. While such precise measurements are unavailable for other program stars, their similarity suggests the same scales might reasonably be inferred to apply. 

The Alfv\'en radius $R_{\rm A}$ is the point at which the magnetic and kinetic energy densities in the wind are equal, and is typically taken as the outer boundary of the magnetic confinement region. At the magnetic equator $R_{\rm A}$ is related to $\eta_*$ via:

\begin{equation}\label{ralf}
\left[\frac{R_A}{R_*}\right]^{2q - 2} - \left[\frac{R_A}{R_*}\right]^{2q - 3} = \eta_*
\end{equation}

\noindent where for a dipole $q=3$ \citep{ud2002}. To achieve Alfv\'en radii of $2.5 R_*$, as implied by the radii of perturbations in $\alpha$ Cyg's wind, $\eta_*\sim 25$ is required. The magnetic field strengths necessary for $\eta_{*} \le 25$ assuming theoretical mass loss rates are indicated in Fig. \ref{bd_lim} with the dotted lines for theoretical \mdot, while dot-dashed lines correspond to observed \mdot. The probabilities P($\eta_*\le 25$) for both theoretical and observed \mdot~are given in Table \ref{eta_bd_lim}. With theoretical mass-loss rates, $\eta_*\le25$ with greater than 95.4\% credibility for all stars; with observed \mdot, $\eta_*\le25$ with greater than 95.4\% credibility for all stars but $\beta$ Ori.

\begin{table}
\centering
\caption[Dipolar upper limits]{95.4\% credible upper limits for $B_d$, the probabilties that that $\eta_*\le X$, and the odds ratios $\log{(M_0/M_1)}$ for Stokes $V$ and diagnostic null.}

\resizebox{8.5 cm}{!}{
\begin{tabular}{|l|rrrrrr}
\hline
\hline
	&	$\beta$ Ori 	&	4 Lac 	&	$\eta$ Leo 	&	HR1040 	&	$\alpha$ Cyg 	&	$\nu$ Cep 	\\
\hline									
			
$B_{\rm d}(P_{\rm cred} = 0.954)$ (G)	&	22	&	14	&	15	&	30	&	3	&	13	\\

\hline

\multicolumn{7}{c}{Theoretical \mdot} \\
$P(\eta_*\le 1)$ 	&	0.981	&	0.982	&	0.921	&	0.797	&	0.995	&	0.942	\\
$P(\eta_*\le 25)$ 	&	0.999	&	0.999	&	0.993	&	0.985	&	0.999	&	0.999	\\
\\
\multicolumn{7}{c}{Observed \mdot} \\
$P(\eta_*\le 1)$ 	&	0.216	&	0.818	&	0.798	&	0.609	&	0.895	&	--	\\
$P(\eta_*\le 25)$ 	&	0.929	&	0.992	&	0.976	&	0.961	&	0.995	&	--	\\
\hline													

$\log{(M_0/M_1)}$ (V) & 0.988 & 0.552 & 1.192 & 0.588 & 1.267 & 0.285 \\
$\log{(M_0/M_1)}$ (N) & 0.909 & 0.600 & 1.094 & 0.543 & -0.299 & 0.868 \\


\hline
\hline
\end{tabular}
}
\label{eta_bd_lim}
\end{table}

The spectral variability attributable to the confined winds of known magnetic massive stars is notable for its precise periodicity, regardless of whether the stellar magnetospheres are purely dynamic (slow rotation) or contain a centrifugal component (rapidly rotating) \citep{petit2013}. In many cases, such as the Of?p stars, rotational periods were determined from frequency analysis of spectral and photometric measurements before spectropolarimetry was available for magnetic analysis \citep{how2007, wade2011}. Conversely, as discussed in Section 1, unique periods have never been inferred for $\alpha$ Cyg variables. While there is evidence for rotational modulation of wind features, (e.g. \citealt{k1996a}), this fundamental dissimilarity, in combination with the firm upper limits for $B_d$ and $\eta_*$, strongly indicate that \textit{magnetic wind confinement due to organized magnetic fields is not the cause of the wind variability of BA SG stars}.

\subsection{Local Magnetic Confinement}

Magnetically confined winds as described above are found in the presence of large-scale, organized magnetic fields. However in the case of the Sun by far the strongest magnetic fields are localized in sunspots, which have a substantial impact on the structure of the solar wind. 

Analogous magnetic features have not been detected in early-type stars. However, bright, hot magnetic spots have been conjectured as a consequence of dynamo action within a thin Fe convection zone (FeCZ), arising from an increase in Fe opacity just below the photosphere (\citealt{cant2009}, \citealt{cb2011}). These spots would be transient features, forming when plasmoids rise from the FeCZ, with two footpoints remaining in the photosphere. The magnetic field within the spots is expected to be approximately normal to the stellar surface, with opposite polarities across each pair of spots (assuming magnetic flux conservation). Magnetic flux above the stellar surface could be expected to perturb the wind in its vicinity, potentially leading to a magnetically suspended loop of the sort suggested by \cite{is1997} in their examination of $\beta$ Ori's HVAs. 

Theoretical predictions of the lifetimes, sizes, and number distributions of spots are quite open: spots might last anywhere from hours to decades, and have radii on the order of (or larger than) the pressure scale height $H_p=kT/\mu g$ within the FeCZ. With the temperature boundaries of the FeCZ ($10^5-10^{5.4}$ K), and the mean molecular weights and surface gravities of our sample stars as established through quantitative spectroscopy (\citealt{prz2006}) we find that $H_p$ should be from $\sim$2--8 \rsun~ at the top and bottom of the FeCZ for most stars, although in $\alpha$ Cyg (with its much lower surface gravity) $H_p$ may range up to 30 \rsun.

A generous upper limit to the spot magnetic field can be determined via equipartition of energy within the FeCZ: since the features are likely to expand as they rise due to magnetic buoyancy, the magnetic field should not be stronger than within the FeCZ. A lower limit can then be calculated via simple scaling relations. For BA SGs, magnetic features at the stellar surface could have field strengths between 5 and 1000 G \citep{cb2011}. 

The effect of bright spots, such as those suggested to be associated with magnetic structures such as those described above, on the stellar wind has been investigated by \cite{cran1996}, who found that such spots would induce high-density, low-speed streams due to the enhanced driving force at the base of the wind. The bright spot modifies the wind density $\rho$ and wind velocity $v(r)$ from their unperturbed values $\rho_0$ and $v_0(r)$ \citep{cran1996}:

\begin{figure}
\centering
\includegraphics[width=3in]{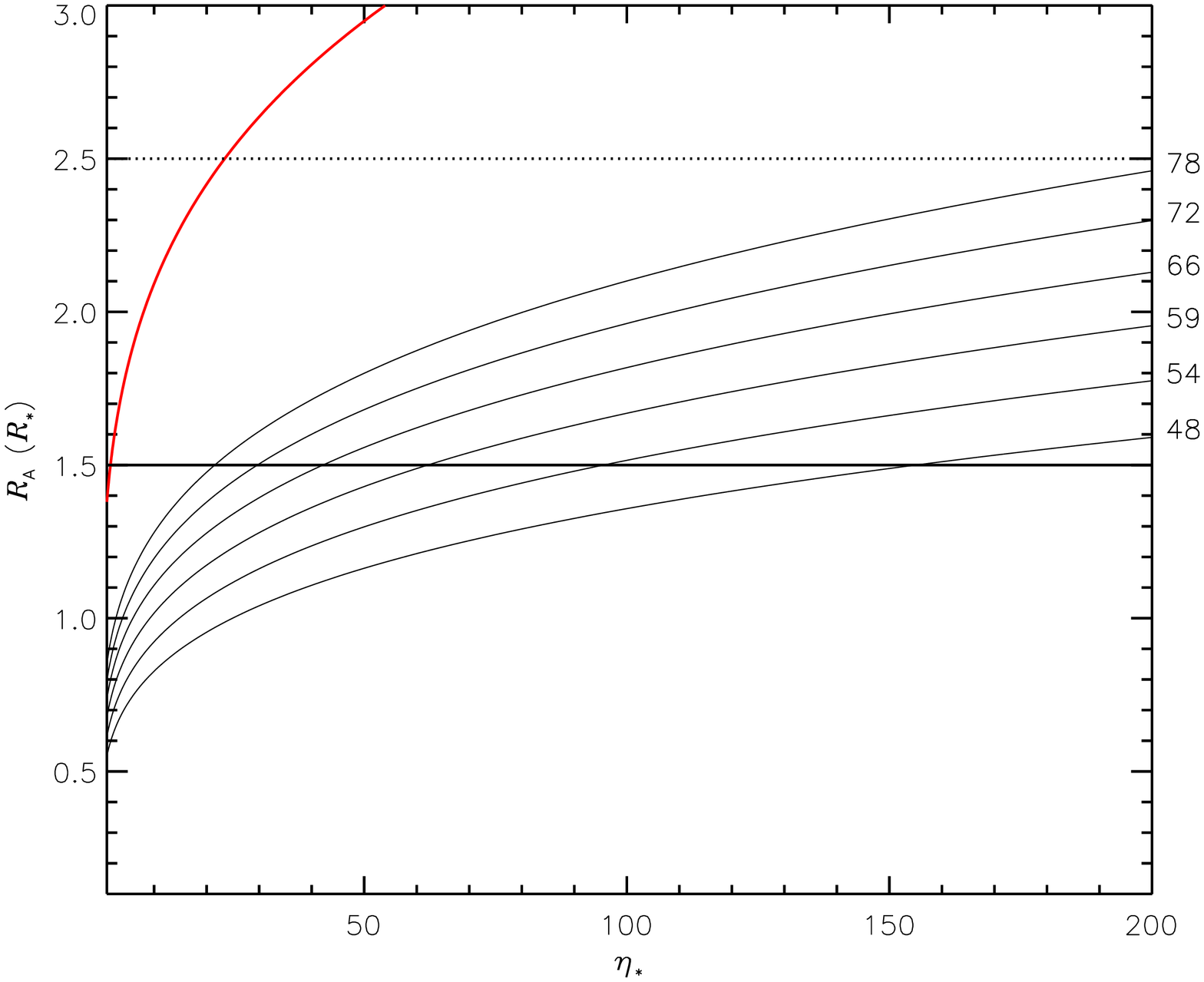} 
\includegraphics[width=3in]{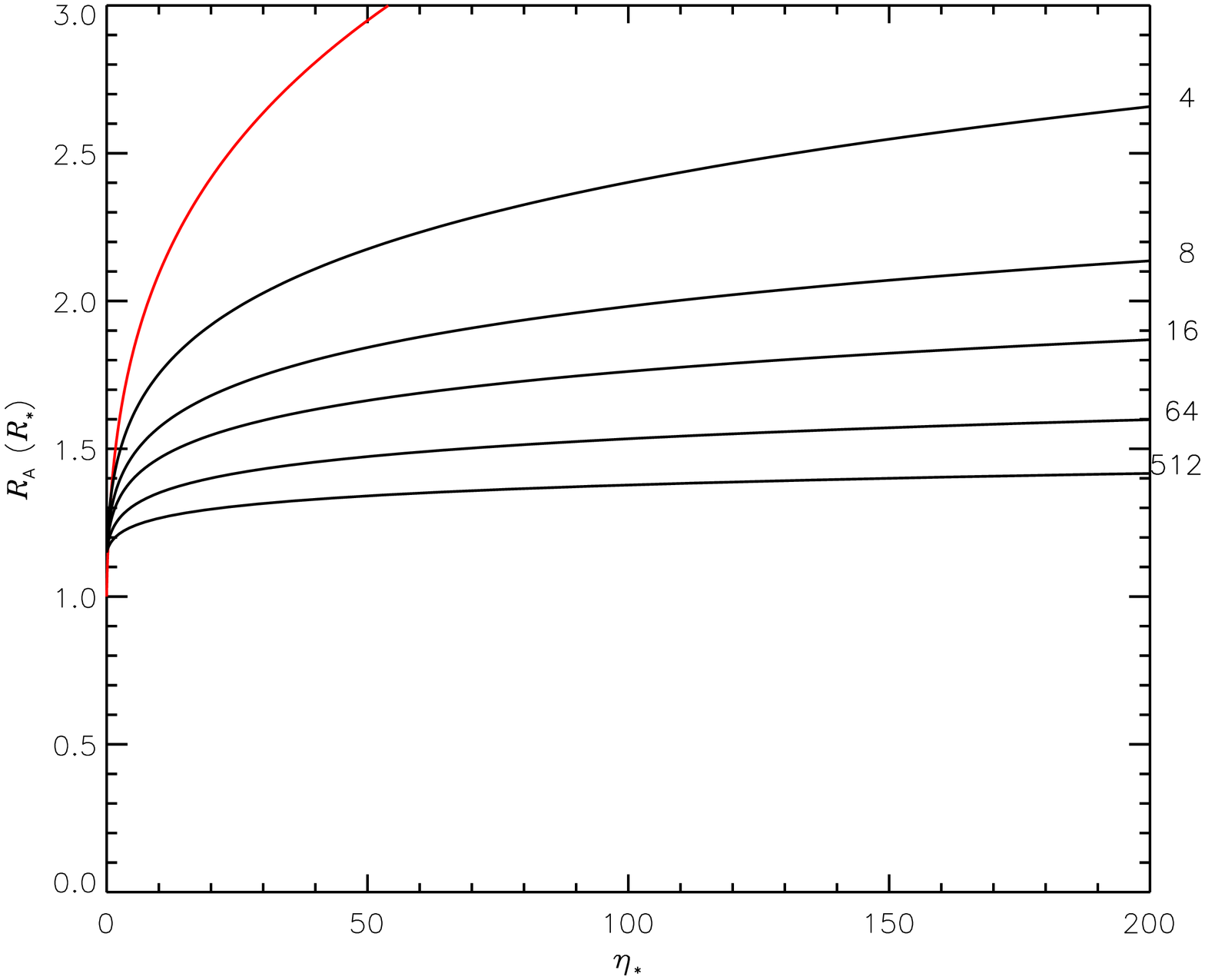}
\caption[Alfven radii of spots]{\textit{Top}: $R_{\rm A}$ as a function of $\eta_*$ (black curves) for different spot separations (annotated, in degrees across the stellar surface). The scaling for a pure dipole is shown in red. The inner and outer regions of interest are shown with solid and dotted lines, respectively. \textit{Bottom}: $R_{\rm A}$ as a function of $\eta_*$ for different exponents $q$ (from top to bottom: 3, 4, 5, 6, 8, 11). Lines are labelled with the number of magnetic poles corresponding to each value of $q$.}
\label{spot_ra}
\end{figure}

\begin{equation}\label{spot_perturb}
\begin{array}{lcl}
\rho & = & \rho_0 (1 + A)^{s/\alpha} \\
v(r) & = & v_0(r) (1 + A)^{(1 - s)/\alpha}
\end{array}
\end{equation}

\noindent where $A = (T_{S}^4 - T_{D}^4)/T_{D}^4 > 0$ is a dimensionless constant relating the relative effective temperatures of the spot, $T_S$, to the disk, $T_D$; $\alpha = 0.60$ is the CAK line-driving constant; and $s = 1.77$ is an exponential constant found through simultaneous constraint of the two fitting functions in Eq. \ref{spot_perturb} via conservation of mass for a spot with a gaussian brightness distribution. From the definition of the wind magnetic confinement parameter \citep{ud2002}, $\eta(r,\theta)$ is modified above the bright spot from its unperturbed value $\eta_0(r,\theta)$ such that

\begin{equation}\label{eta_spot}
\eta_{{\rm S}}(r,\theta) = \frac{\eta_0(r,\theta)}{(1 + A)^{(2 - s)/\alpha}}
\end{equation}

The effect is to reduce $\eta(r,\theta)$, due to the overall enhancement of the wind momentum, with the magnitude of reduction increasing with $A$. Assuming $A = 0.5$ (implying temperature differences of $\sim$1000 K), as used by \cite{cran1996}, $\eta(r,\theta)$ should be depressed by around 10--15\%. In the context of the \citeauthor{cb2011} model, the spot's brightness is a result of the magnetic field (which results in a lower density that makes deeper, hotter parts of the atmosphere visible; see David-Uraz et al., in prep.), so the assumption of a brighter spot in practice requires a stronger magnetic field (and vice versa). The detailed atmospheric modeling required to explore the precise relationship between magnetic field strength and spot brightness is outside the scope of this paper; in any case, spot brightness may play a limiting role in the magnetic field's ability to confine the wind. 

Since the extent of magnetic field lines is partly a function of the separation of the magnetic poles, the separation of the spots on the stellar disk also has an effect on $R_{\rm A}$. If we approximate the spots as the two poles of a magnetic dipole, $R_{\rm A}$ for a given $R_*$ should satisfy Eq. \ref{ralf}, with the modification that the stellar radius $R_*$ is replaced by the spot separation $D_{\rm spot} = R_*\sin{\theta/2}$, where $\theta$ is the angular separation of the spots. Eq. \ref{ralf} then becomes:

\begin{equation}\label{raspot}
\left[\frac{R_A}{R_*\sin{\theta/2}}\right]^{2q - 2} - \left[\frac{R_A}{R_*\sin{\theta/2}}\right]^{2q - 3} = \eta_*
\end{equation}

This increases the value of $\eta_*$ in Eq. \ref{ralf} required to achieve a given $R_{\rm A}$, as shown in Fig. \ref{spot_ra}. To investigate the plausibility of Eq. \ref{raspot}, we examine a typical solar coronal giant arch \citep{fab2010}. The solar wind has \mdot~$\sim2-3\times10^{-14}~M_\odot~{\rm yr}^{-1}$ and \vinf~$\sim 400$~\kms. The footpoints of an arch possess $B_{\rm S}~\sim~100$ G, yielding $\eta_*\sim 5\times10^5$ from Eq. \ref{etastar}. With $D_{\rm spot}~\sim~3~\times 10^4~{\rm km}~\sim~0.04~R_\odot~\sim~2.5 ^\circ$, Eq. \ref{raspot} then gives $R_{\rm A}~\sim~0.5$~\rsun. The actual lengths of coronal arches vary from $\sim~0.1~-~1.5$~\rsun, in approximate agreement with Eq. \ref{raspot}.

\cite{is1997} estimated that spots were separated by $\sim75^\circ$, based upon the time between blue- and red-shifted HVA absorption maxima. $R_{\rm A}$ as a function of $\eta_*$ is shown for various separations in the top panel of Fig. \ref{spot_ra}, where $\eta_*$ is depressed by 15\% as per Eq. \ref{eta_spot}. Individual curves are labelled with the spot separation in degrees. Assuming a separation of $\sim75^\circ$, $\eta_*~\sim~25$ should be sufficient to confine the plasma to $\sim1.5 R_*$; plasma can be confined out to $\sim~2.5~R_*$ if $\eta_*~\sim~200$. The surface magnetic field strengths necessary to achieve these values are given in Table \ref{sad-tab-2}.

To investigate the detectability of magnetic spots in high-resolution Stokes $V$ spectra, \cite{kochsud2013} performed detailed spectrum synthesis calculations for a grid of models with \vsini~varying between 10 and 200 \kms, spot angular radii $r_{\rm sp}$ between 2$^\circ$ and 40$^\circ$, and fixed parameters spot magnetic field strength $B_{\rm S}~=~500~{\rm G}$, filling factor $f_{\rm sp}~=~0.5$, \teff = 20 kK, and $i~=~60^\circ$. Each model consisted of a non-magnetic photosphere speckled with a random realization of spots, with equal numbers of positive and negative spots such that the net magnetic flux was zero. \citeauthor{kochsud2013} then calculated LSD Stokes $V$ profiles and \bz, which they used to provide upper limits based on the general characteristics of the MiMeS survey data.

Unsurprisingly, given the complexity of the magnetic topology, \bz~is an insensitive diagnostic of small-scale magnetism. Stokes $V$ is more sensitive: numerical experiments by \citeauthor{kochsud2013} showed that small-scale magnetic fields should register a marginal detection in the Stokes $V$ profile, provided the maximum Stokes $V$ amplitude is at least 5 times the error bar.

The stars considered in our sample differ considerably from the template used by \citeauthor{kochsud2013}, in particular being much cooler. \citeauthor{kochsud2013} note that decreasing \teff~to 10 kK will increase the Stokes $V$ amplitude by $\sim$30\% for slow rotators. Furthermore, the median error bars in the current sample are substantially better than those in the MiMeS survey data (25 G, as compared to 4.3 G for these stars). 

\begin{figure*}
\centering
\begin{tabular}{ccc}
\includegraphics[width=2.25in]{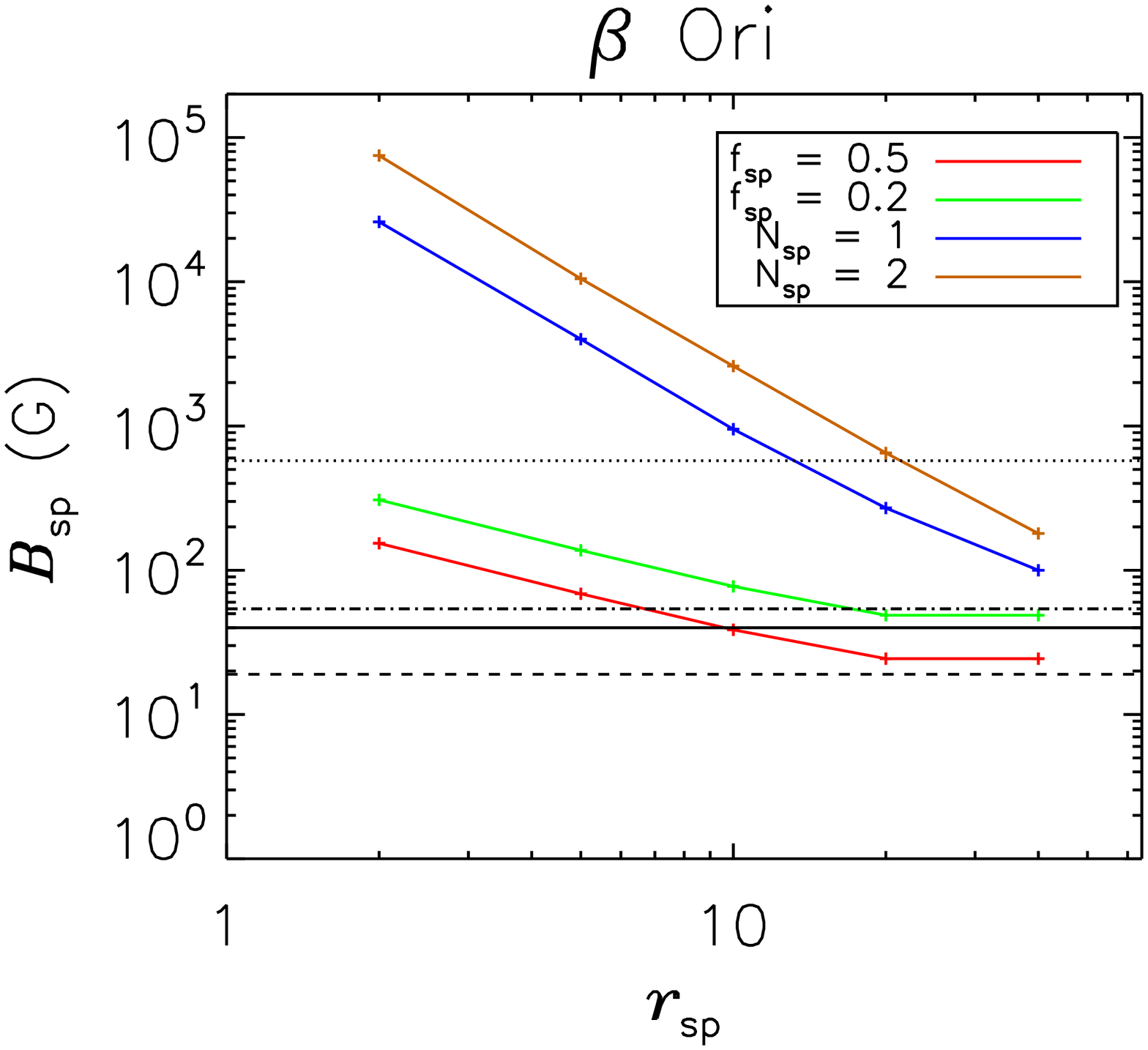} &
\includegraphics[width=2.25in]{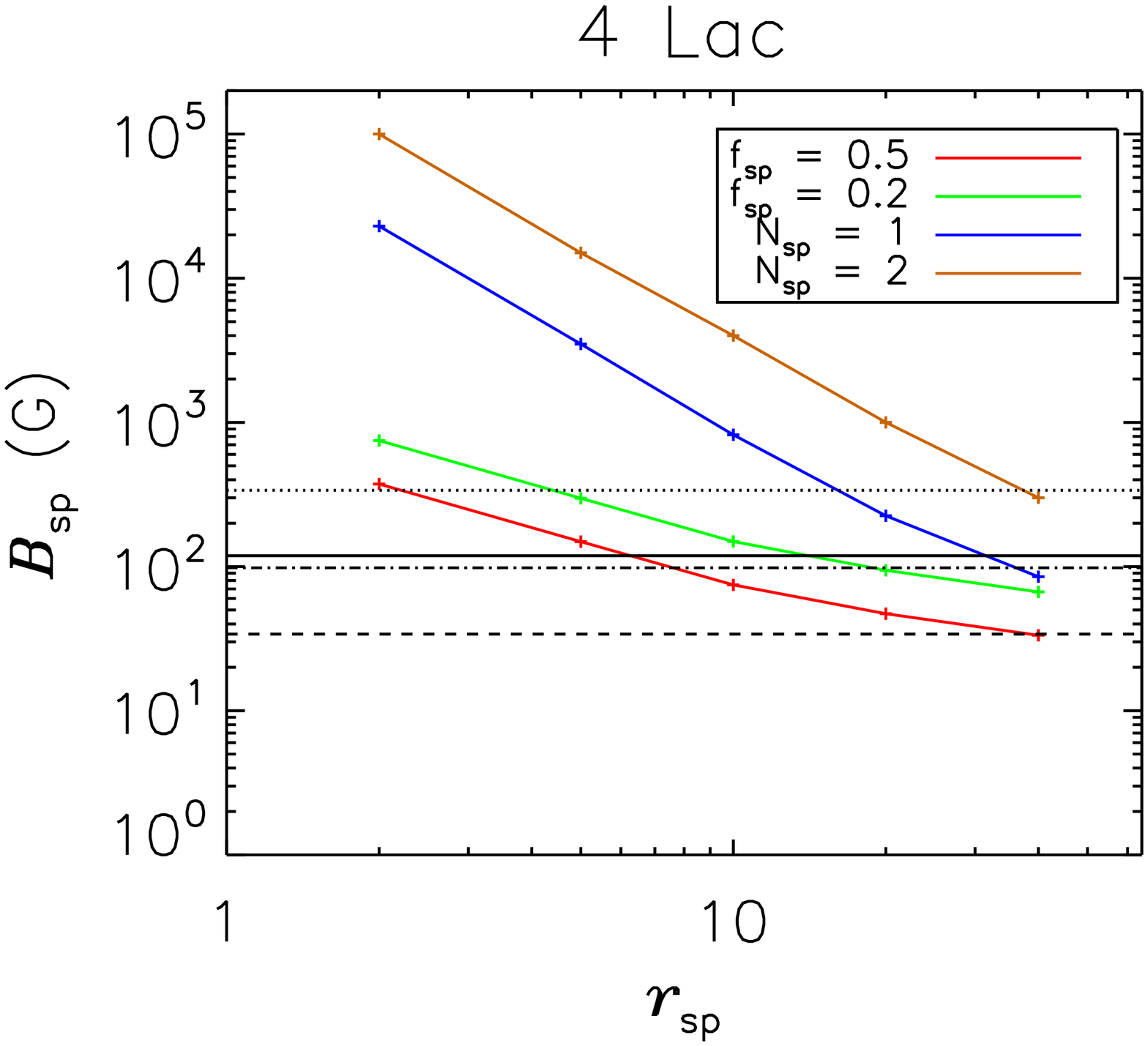} &
\includegraphics[width=2.25in]{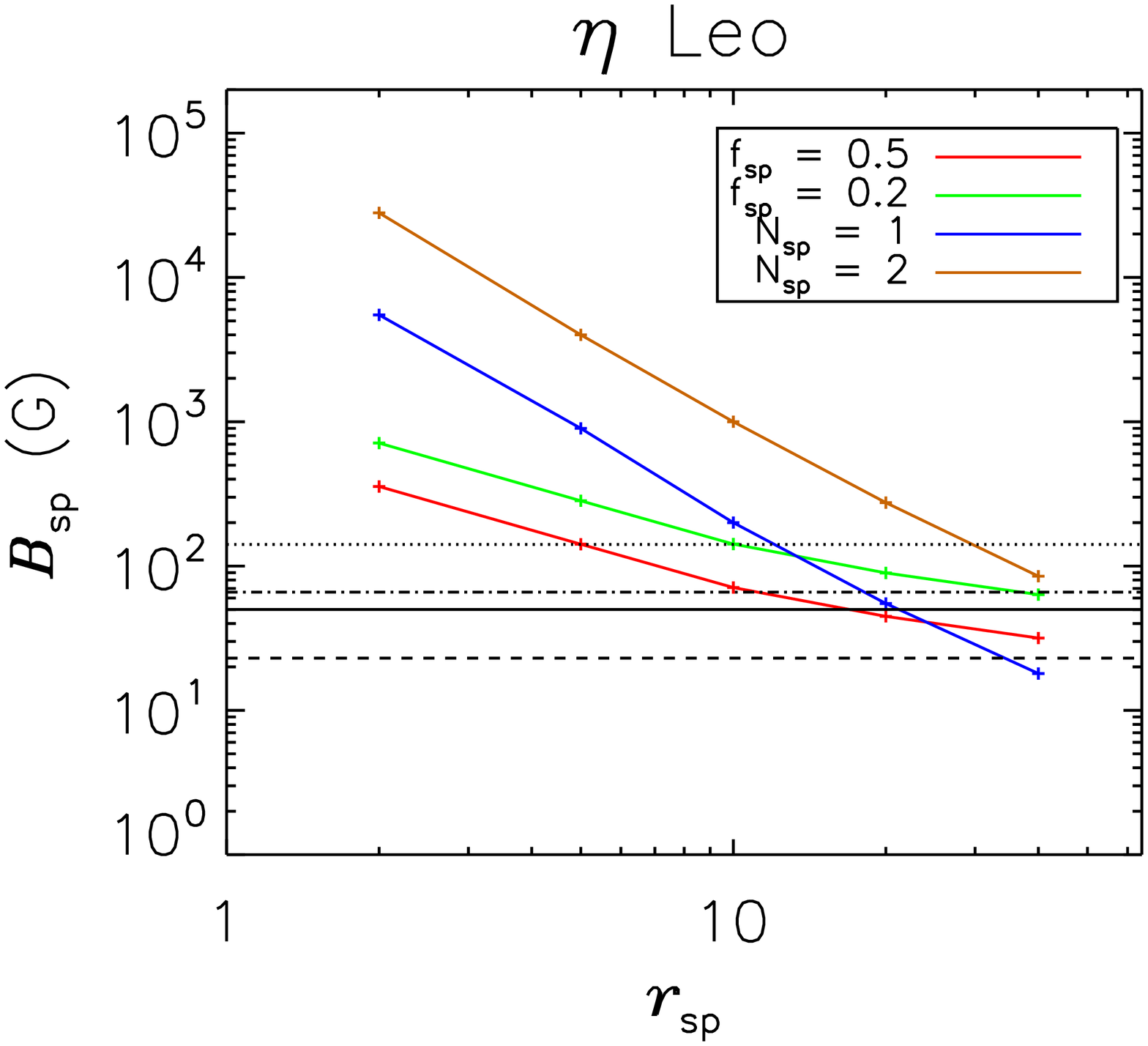} \\
\includegraphics[width=2.25in]{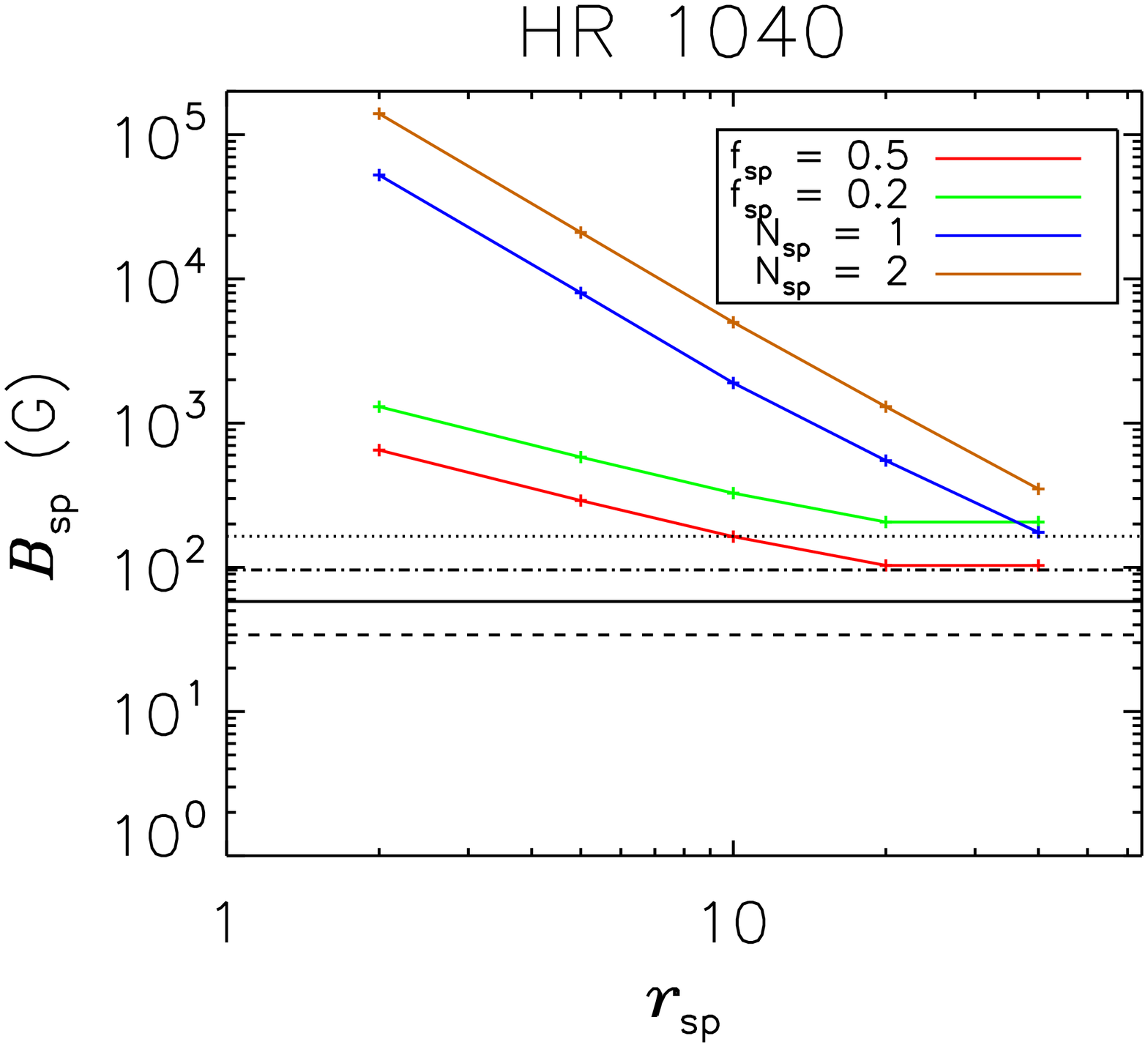} &
\includegraphics[width=2.25in]{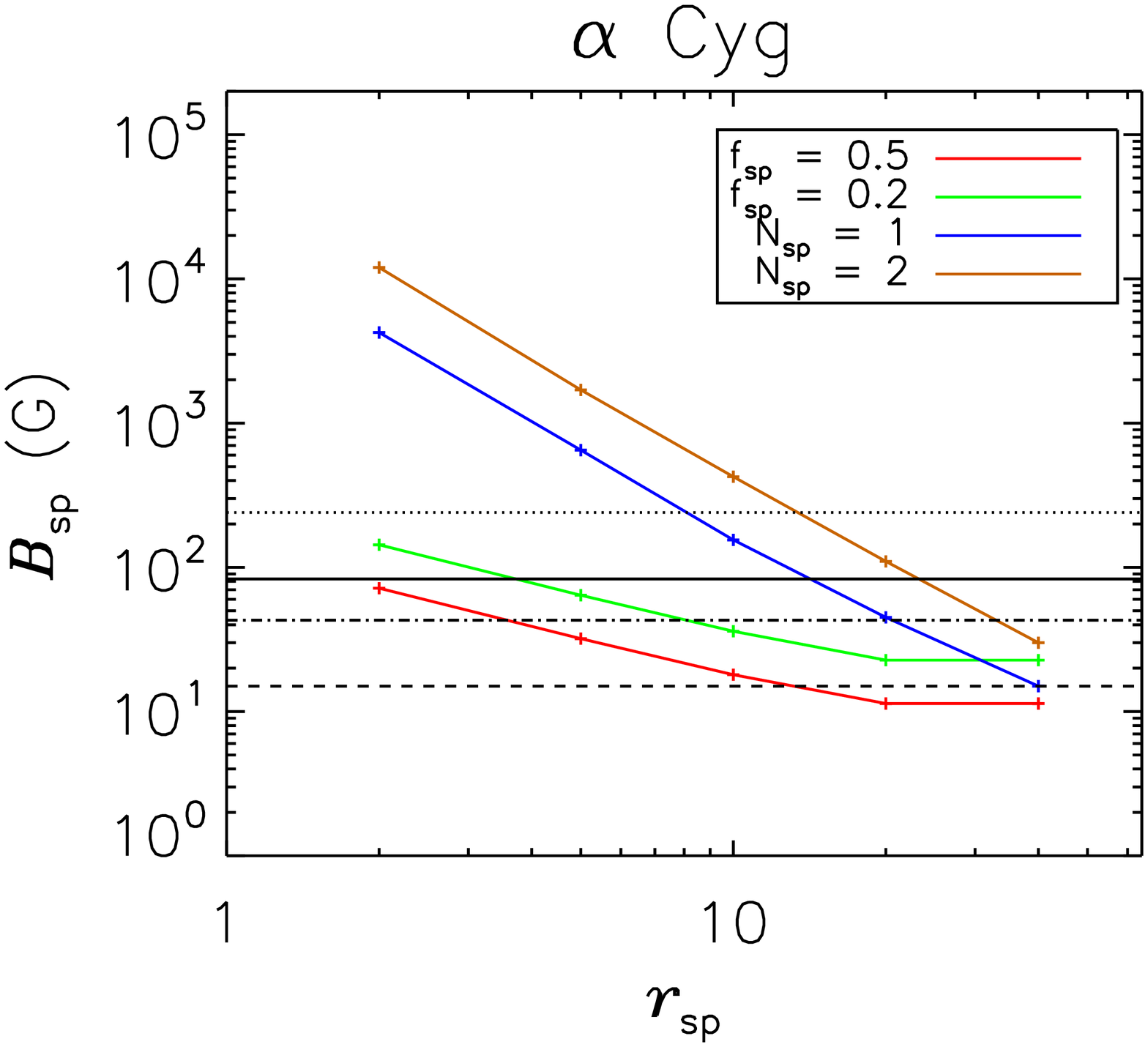} &
\includegraphics[width=2.25in]{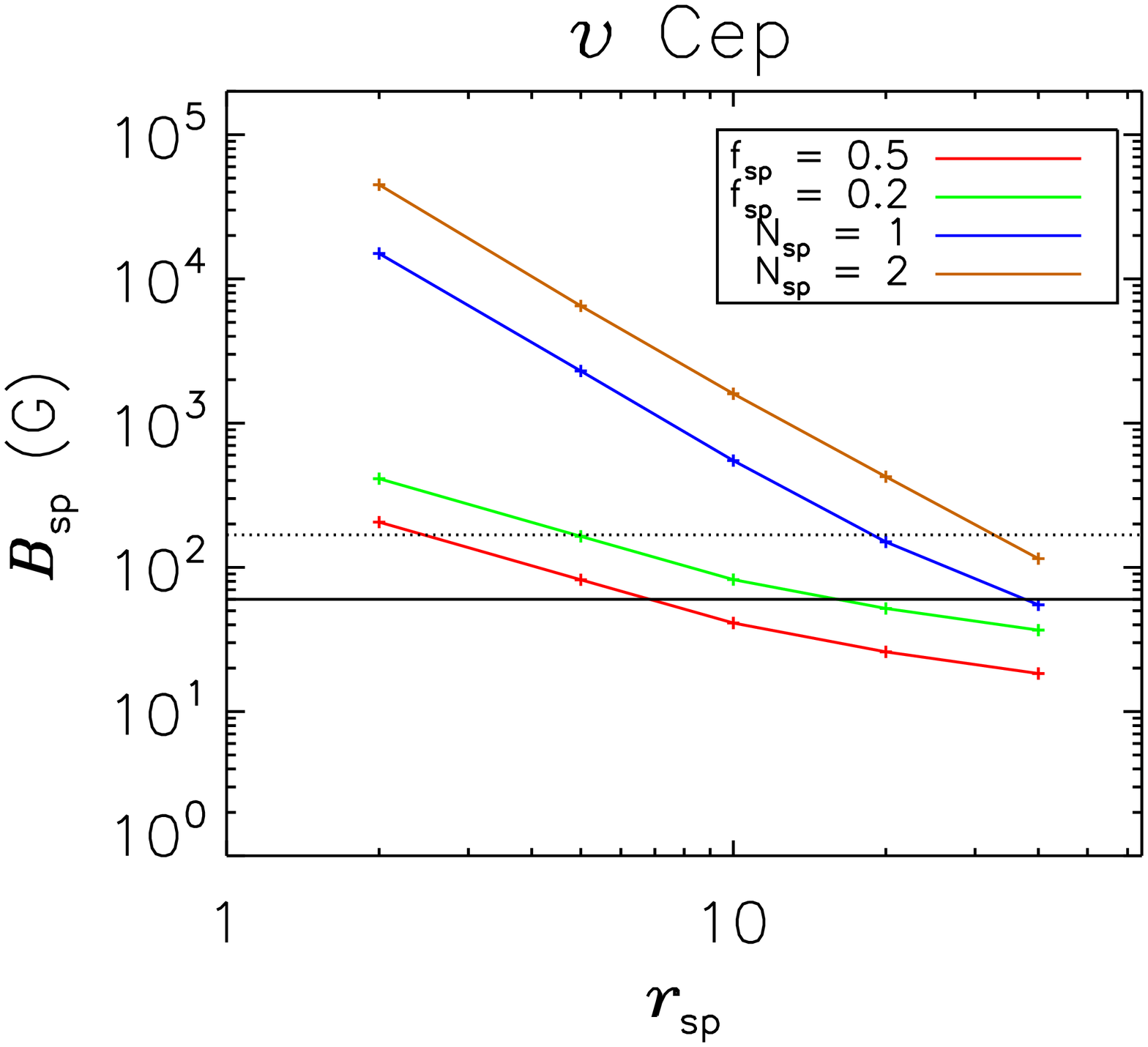} \\
\end{tabular}
\caption[Spotted field upper limits]{Upper limits on spot surface fields $B_{\rm sp}$ as a function of spot radius $r_{\rm sp}$ for the highest SNR observations for each star, for $f_{\rm sp} =0.5$ (red), $f_{\rm sp} =0.2$ (green), $N_{\rm sp}=1$ (blue), and $N_{\rm sp}=2$ (brown). Solid lines indicate $B_{\rm sp}(\eta_* = 25)$, dotted lines $B_{\rm sp}(\eta_* = 200)$, assuming two spots separated by 75$^\circ$ and theoretical \mdot. Dashed and dot-dashed lines indicate the same limits for measured \mdot.}
\label{spot_lim}
\end{figure*}

After scaling for \teff, we find approximate upper limits for spot magnetic fields by computing the difference between the expected maximum Stokes $V$ amplitude for a 500 G spot field (Fig. 5 of \citealt{kochsud2013}), and five times the Stokes $V$ error bar in the LSD measurements (the minimum amplitude to register a marginal detection). The targets 4 Lac, $\eta$ Leo, and $\nu$ Cep, which are slow rotators, were compared to the 10 \kms~\vsini~curve; the other stars were compared to the 20 \kms~curve. The resulting upper limits are indicated as red curves in Fig. \ref{spot_lim} (compare with Fig. 7 of \citealt{kochsud2013}). For large spots, the upper limits in Fig. \ref{spot_lim} approach those obtained for the dipolar magnetic field in Fig. \ref{bd_lim}.

These calculations assume a heavily spotted surface ($f_{\rm sp}=0.5$). Decreasing $f_{\rm sp}$ will of course decrease both Stokes $V$ and \bz, however the effect is not linear as with fewer spots the cancellation of magnetic flux becomes less efficient, e.g. a factor 2.5 reduction in $f_{\rm sp}$ will decrease the Stokes $V$ amplitude by only 50\%. Predictions for $f_{\rm sp}=0.2$ are indicated as green curves in Fig. \ref{spot_lim}. The best constraints are obtained for $\alpha$ Cyg, for which even small spots with $B_{\rm S} > 150$ G can be ruled out; in general, $B_{\rm S} < 10^2 - 10^3$ G, depending on spot size.

Large filling factors are unlikely to be compatible with extended magnetically confined wind structures, since the distance between any two given spots of opposite polarity (the foot-points of magnetic loops) will in general be small if the spots are randomly distributed on the photosphere. Values of $\eta_*$ calculated assuming a dipolar magnetic field must be modified by replacing the exponent $q=3$ in Eq. \ref{raspot} with a larger value, e.g., $q=4$ for a quadrupole, $q=5$ for an octopole, etc. In the limit of many spots, the photospheric magnetic field can be approximated with an extremely high-order multipole and a correspondingly large value of $q$, e.g. a photosphere with $f_{\rm sp}~=~0.5$ and $r_{\rm sp}~=~2^\circ$ will have $\sim$1640 magnetic poles, in which case, $\eta_*$ of several thousand would be required to confine the wind even to 1.5 $R_*$. The effect of varying $q$ is shown in the bottom panel of Fig. \ref{spot_ra}.

\begin{figure*}
\centering
\begin{tabular}{ccccc}
\includegraphics[width=1.25in]{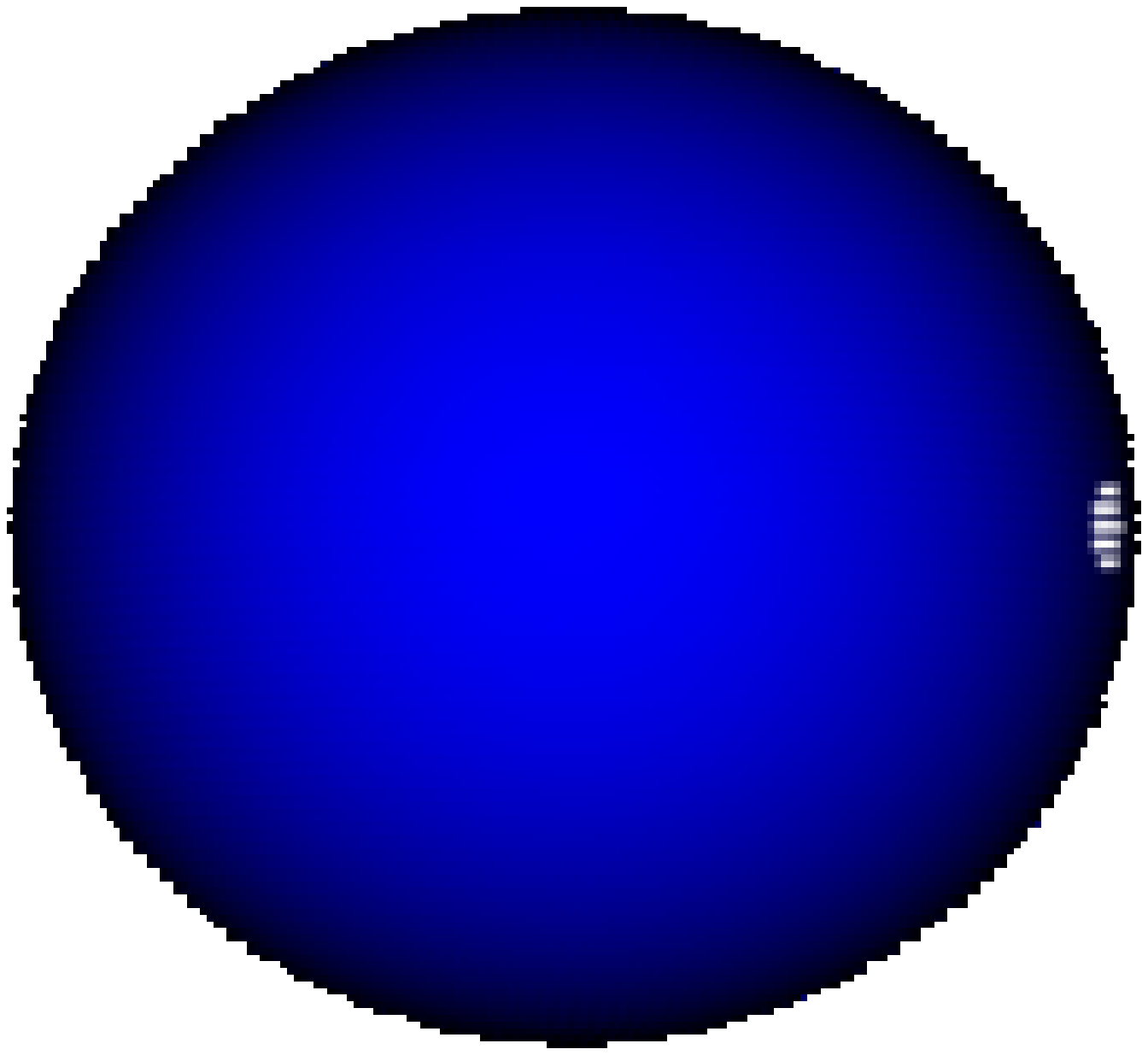} &
\includegraphics[width=1.25in]{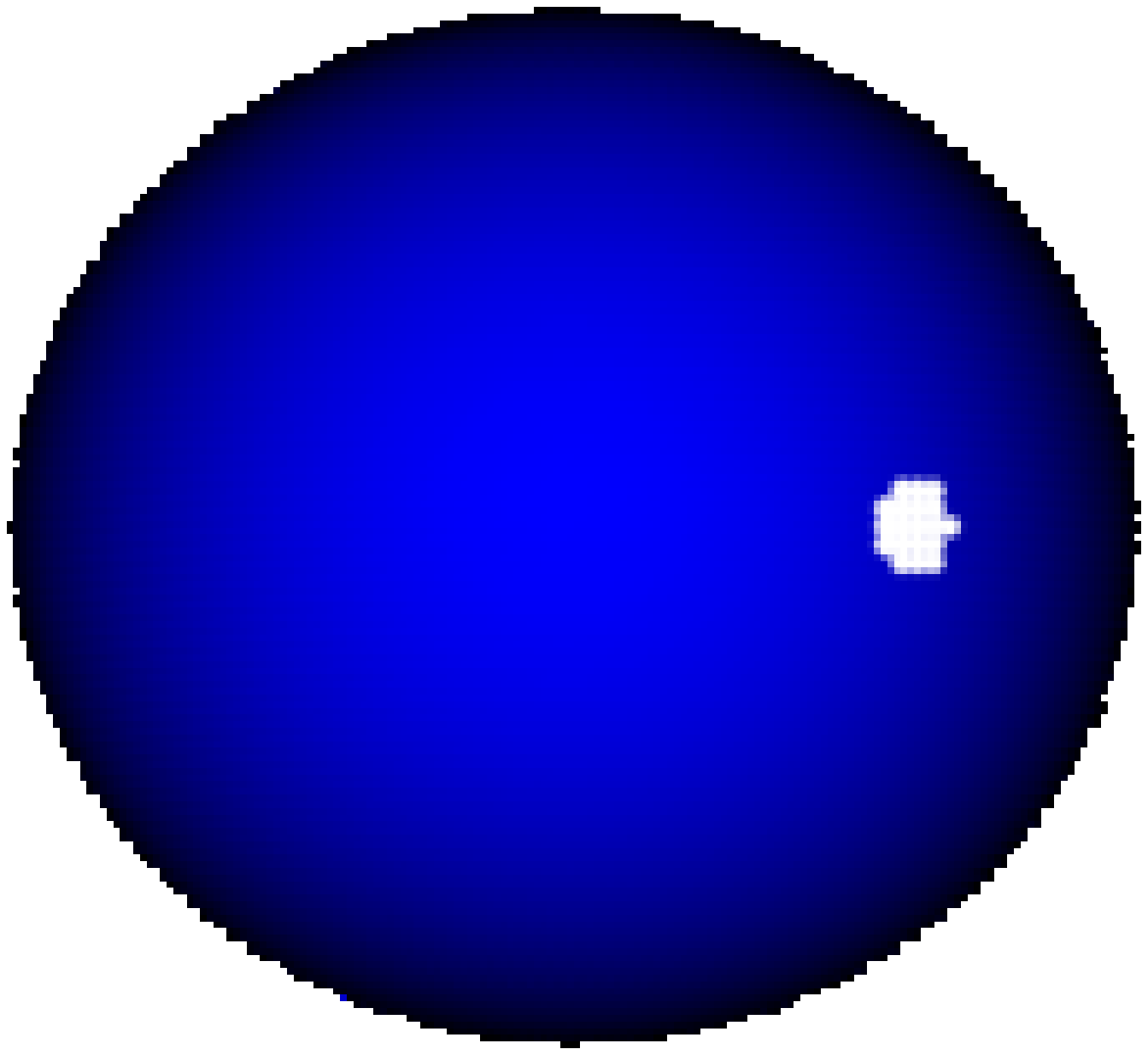} &
\includegraphics[width=1.25in]{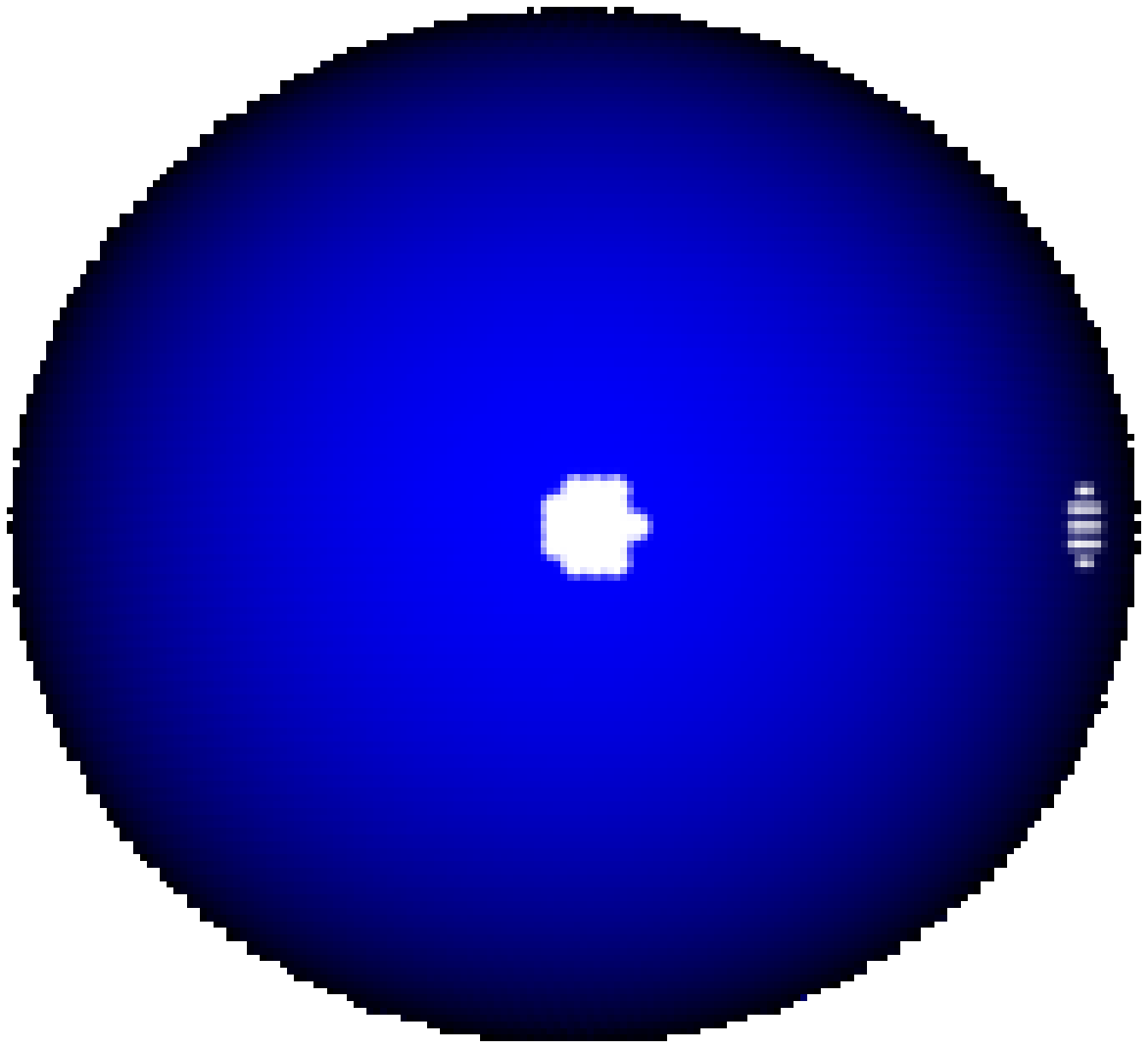} &
\includegraphics[width=1.25in]{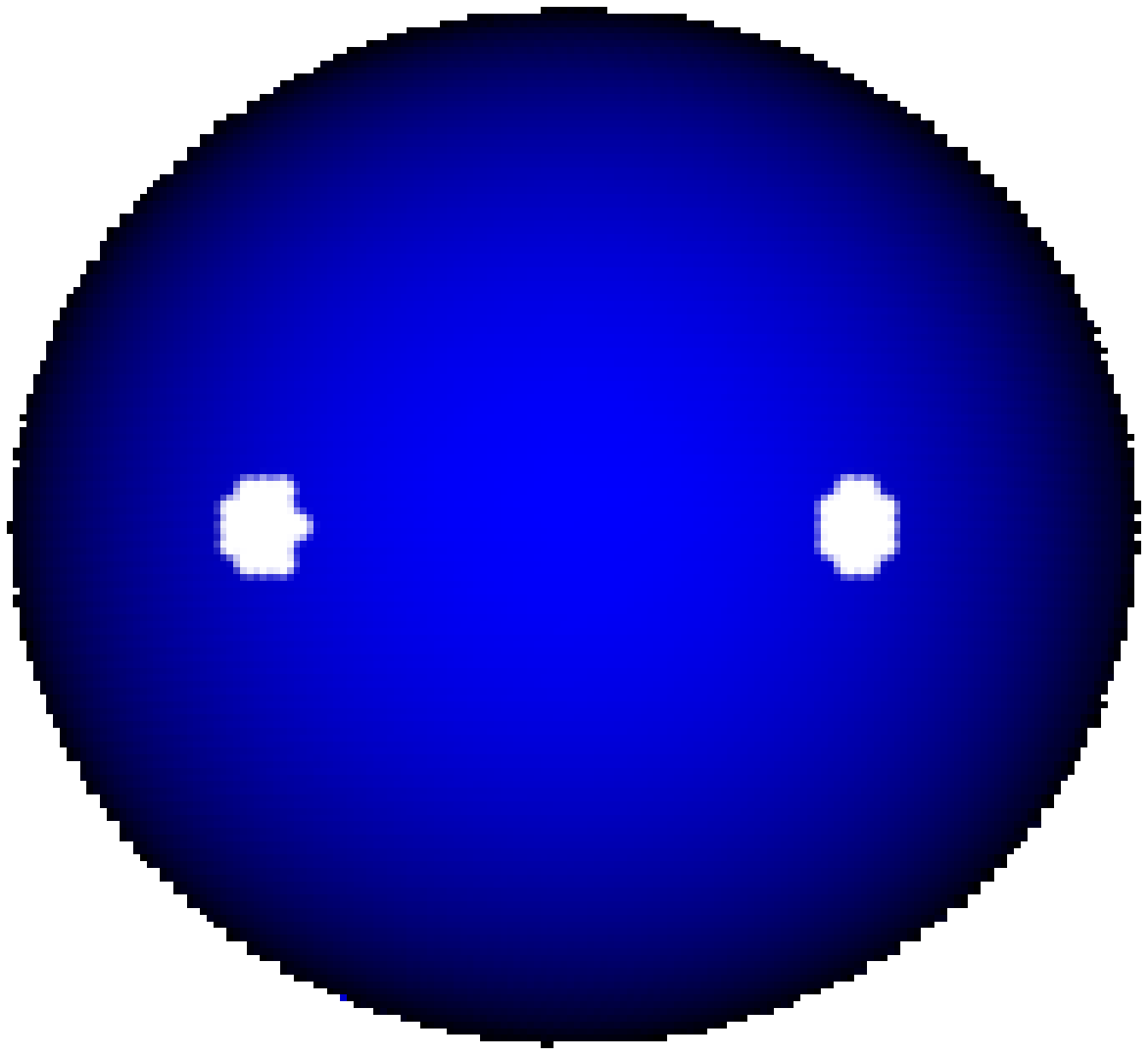} &
\includegraphics[width=1.25in]{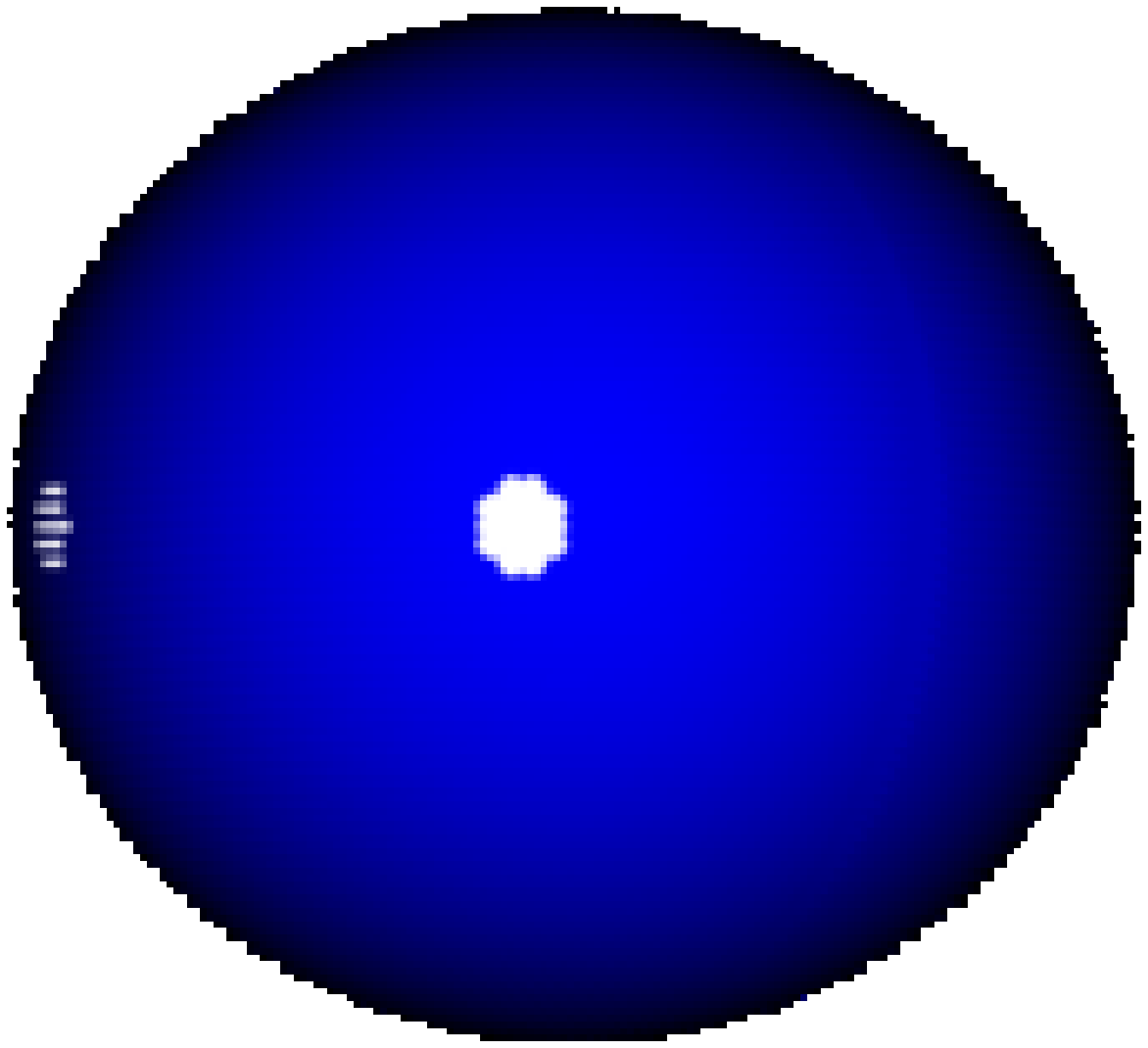} \\
\includegraphics[width=1.25in]{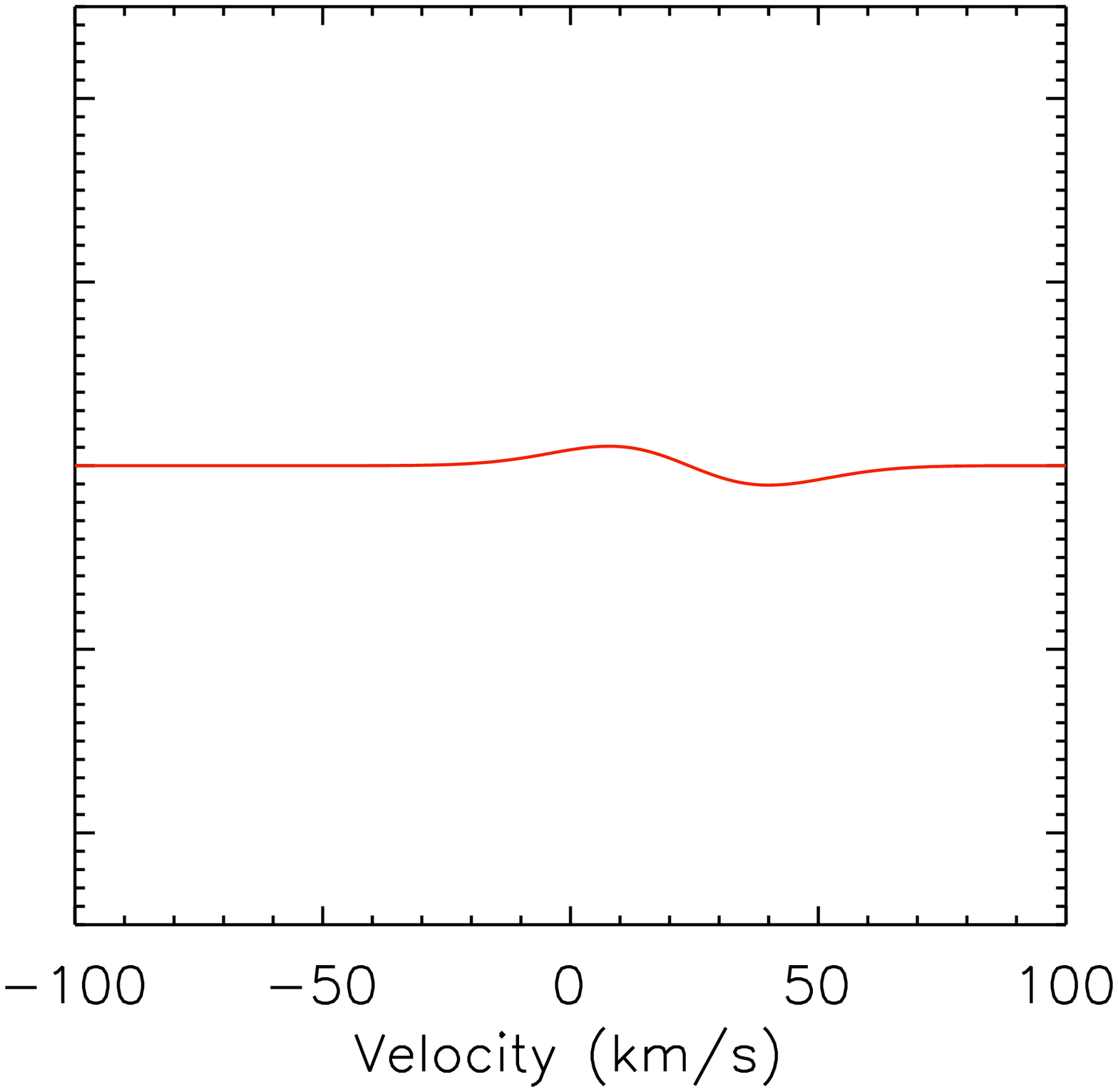} &

\includegraphics[width=1.25in]{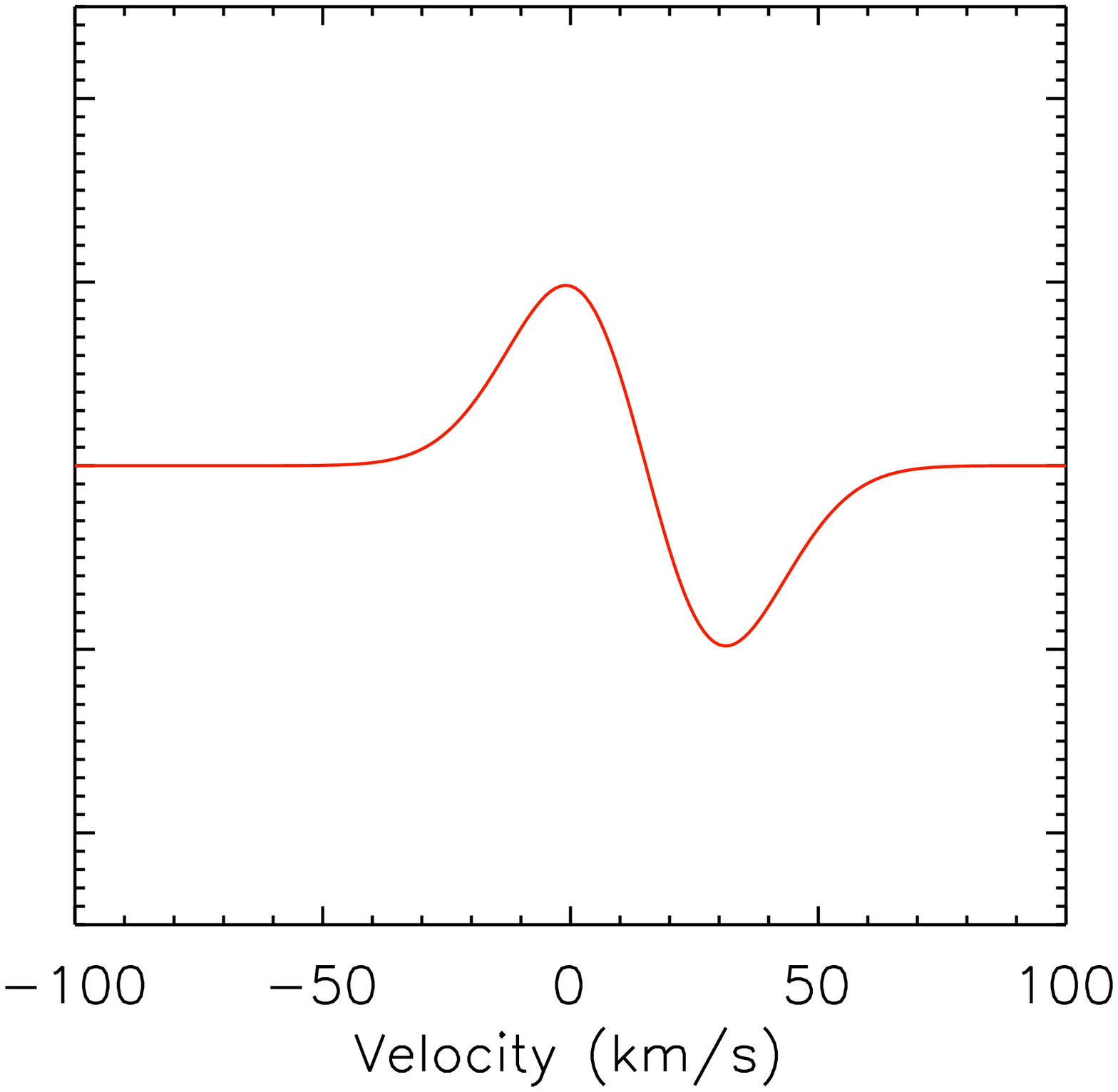} &

\includegraphics[width=1.25in]{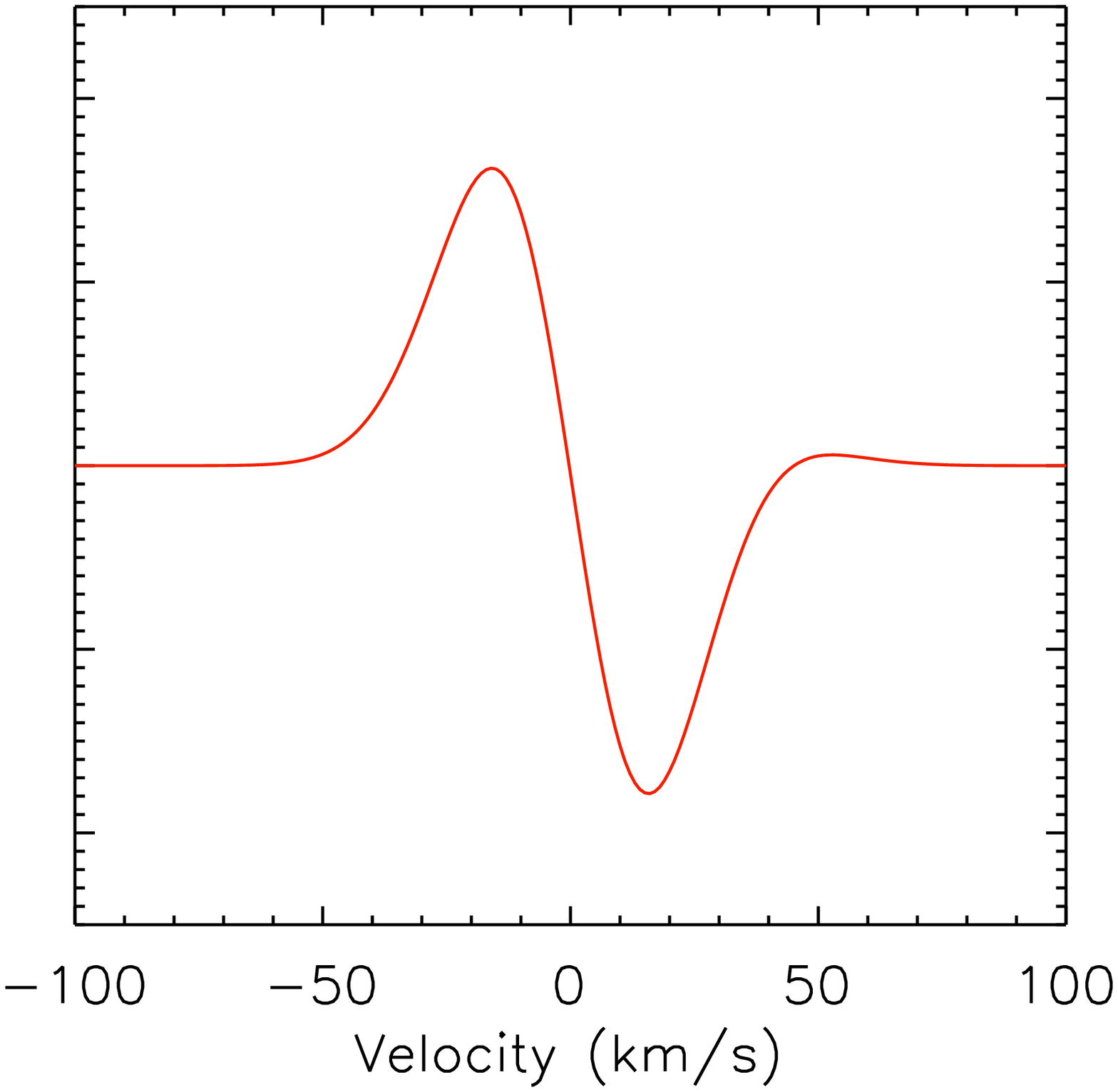} &

\includegraphics[width=1.25in]{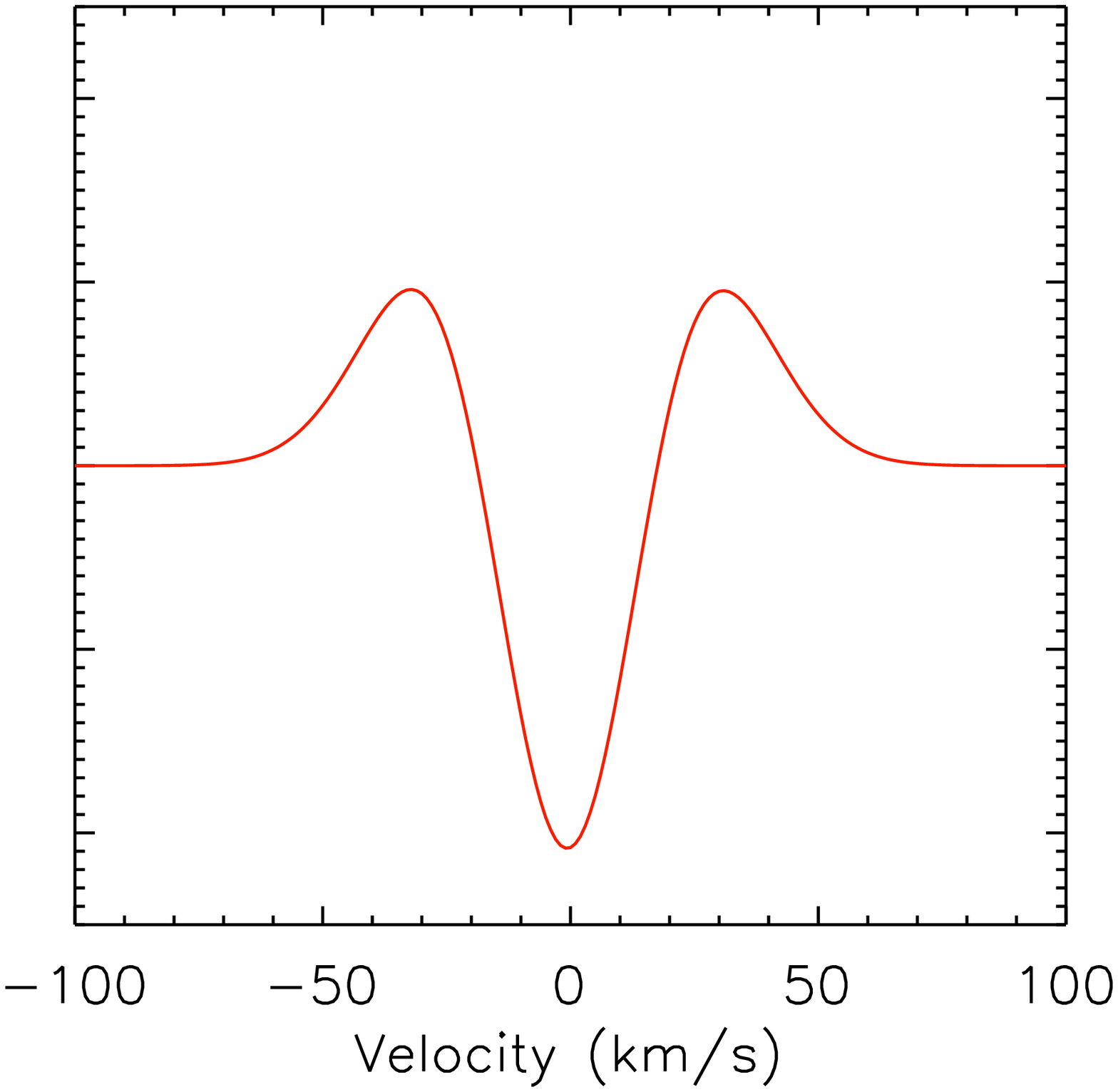} &

\includegraphics[width=1.25in]{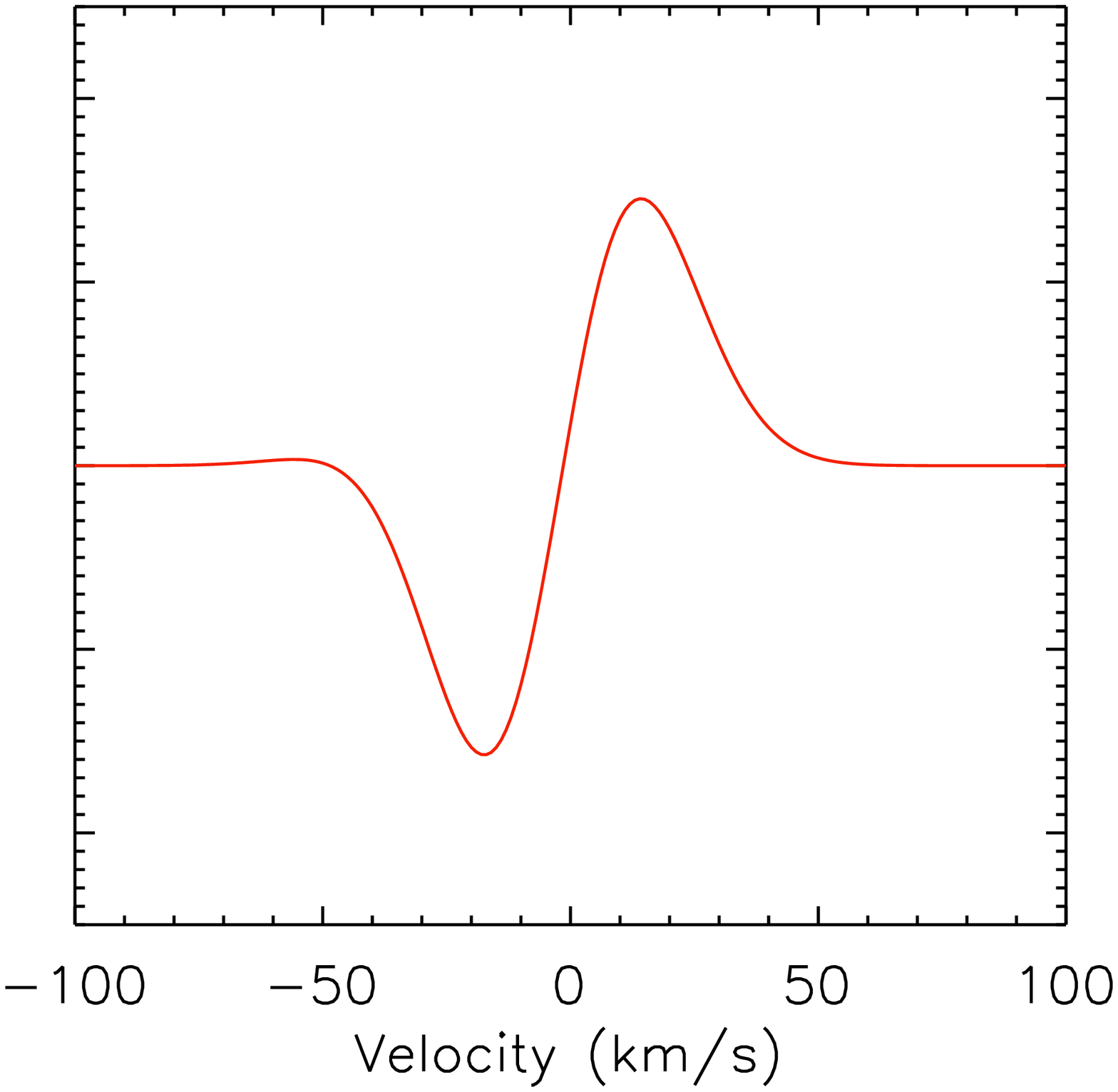} \\
\end{tabular}
\caption[Two-spot Stokes $V$ variation]{Expected variation in Stokes $V$ for two magnetic spots separated by 75$^\circ$, in rotational phase increments of 0.1. The amplitude of the signal is strongest when only a single spot is visible; furthermore, for the majority of rotational cycle, Stokes $V$ is dominated by a single spot.}
\label{spot_model}
\end{figure*}

It is much easier to confine the wind if the magnetic flux is dominated by a single spot pair. A simple model was constructed to investigate the Zeeman signatures that would be produced by such a confinguration (see Fig. \ref{spot_model}). The field within the spots is normal to the stellar surface, with no magnetic field elsewhere; the surface field strength $B_S$ falls off azimuthally from the centre of the spot as $e^{-(r/r_{spot})^2}$, roughly reflecting the field structure observed in sunspots. Spots are of identical $|B_S|$, opposite polarity, and identical angular radii (between $2^\circ$ and $40^\circ$ of the stellar surface, following \citeauthor{kochsud2013}). Limits were found by calculating the surface field strength necessary for the peak-to-peak amplitude of the Stokes $V$ signature to equal 5 times the noise level.

Given the relatively large inferred separation between spots ($\sim75^\circ$, as discussed above), at any given time it is possible that only a single spot will be visible. The resulting limits are shown in Fig. \ref{spot_lim} for the case of a single visible spot (blue lines) and two visible spots (brown lines). Note that, due to flux cancellation, a single spot is typically easier to detect than a pair of spots.

The magnetic field strengths necessary to obtain $\eta_*=25$ and $\eta_*=200$, calculated from Eqs. \ref{eta_spot} and \ref{raspot} with theoretical \mdot, are indicated with solid and dotted lines, respectively; dashed and dot-dashed lines indicate field strengths for the same values of $\eta_*$ calculated using observed values of \mdot. Assuming theoretical mass-loss rates, for all stars, a single spot-pair with a magnetic field strong enough to confine the wind out to 2.5 $R_*$ could have remained undetected, so long as $r_{\rm sp} < 10^\circ$. With measured \mdot, in only a few cases can even the largest spots be ruled out.

As discussed in Section 3, some observations were fortuitously obtained during HVA events (see Fig. \ref{Halpha_profs} and Table \ref{btab-mean}). In the case of $\alpha$ Cyg, the SNR is too low to obtain meaningful constraints. The same is true for the final observation of $\beta$ Ori, during which a magnetic spot may have been visible on the limb. In the case of $\nu$ Cep, however, the strongest HVAs also correspond to the highest SNR LSD profiles. Hence, the constraints provided in Fig. \ref{spot_lim} bear directly on the magnetic hypothesis for HVA events, and it can be concluded that if magnetic spots play an important role, both $r_{\rm sp}$ and $f_{\rm sp}$ must be relatively small. However, it should be remembered that as this conclusion is based upon a theoretical mass-loss rate, measurement of \mdot~for this star would likely revise the magnetic field strengths necessary for wind confinement downward by a factor of 5 to 10, as for the other stars in this sample, in which case these upper limits would no longer be able to rule out spots of even the largest size.

\section{Conclusions}

We have observed 6 BA SG stars using the high-dispersion spectropolarimeters ESPaDOnS and Narval. Spectral analysis of their \halp~lines reveals wind behaviour consistent with previous observing campaigns. HVAs are detected in the \halp~lines of $\alpha$ Cyg and $\nu$ Cep, while weak HVA-like behaviour is observed in $\beta$ Ori's \halp~time series. No magnetic field is detected in any observation, with no Zeeman signatures visible in the LSD Stokes $V$ profiles, and a mean precision of $\langle \sigma_{B_Z} \rangle = 4$ G in the longitudinal field, with the highest SNR LSD profile yielding $\sigma_{B_Z} = 0.5$ G. 

Spectral modeling of the LSD profiles, in conjunction with the Bayesian approach of \cite{petit2012a}, yields upper limits on the dipolar field strength that effectively rule out magnetic wind confinement due to a dipolar magnetic field with at least 95.4\% confidence for all stars but $\beta$ Ori, which can only be constrained to 92.9\% confidence if empirical mass-loss rates are used. These measurements, and their resulting upper limits, represent the tightest constraints yet established for the magnetic fields of BA SG stars. However, these upper limits are sensitive to the mass-loss rates adopted, and should be revised if more accurate models of non-spherically symmetric wind structures, enabling more accurate measurement of \mdot, are developed.

If magnetic confinement of the stellar wind is a significant factor in the circumstellar dynamics of BA SG stars, the underlying field structure must be more complex than a global dipole. The existence of bright magnetic spots cannot be ruled out entirely based on these observations. However, a consideration of the effects of bright magnetic spots on the stellar wind, taking into account both the line-driving force and the reduction in the Alfv\'en radius due to spot separation, requires that the spot surface magnetic fields be one to two orders of magnitude stronger than the lower limits established by \cite{cb2011} if they are to confine the wind out to the radii inferred via spectro-interferometry \citep{ches2010, k2012}. Whether it is realistic to produce such strong fields in the subsurface convection zones of these stars remains to be investigated. Furthermore, if theoretical mass-loss rates are assumed the highest SNR observations of $\alpha$ Cyg, $\beta$ Ori, and $\nu$ Cep are able to effectively rule out magnetic wind confinement out to the observationally inferred radii of wind perturbations, even for very small spots, unless the spots cover substantially less than 20\% of the surface.

Due to the unknown sizes and lifetimes of spots, and the current inability to predict the circumstellar dynamics of $\alpha$ Cyg variables, magnetic detection would require a significant investment in telescope time: first, spectral monitoring to identify moments of maximum spot detectability (e.g. the onset of an HVA event); second, deep spectropolarimetric observations in order to obtain LSD profiles with a SNR sufficient to detect spot magnetic fields. Such a program would benefit from 3D MHD modeling, which would help to investigate the feasibility of the bright magnetic spot model as the origin of HVA events, as well as from the development of non-spherically symmetric mass-loss models.

\section*{Acknowledgments}

G.A.W. and D.A.H. acknowledge Discovery Grant support from the Natural Sciences and Engineering Research Council (NSERC) of Canada. J.H.G. acknowledges former support from an Alexander Graham Bell Canada Graduate scholarship from NSERC. This research used the facilities of the Canadian Astronomy Data Centre operated by the National Research Council of Canada with the support of the Canadian Space Agency. M.E.S. would like to thank James Silvester, Thomas Rivinius, Nancy Morrison, Olivier Chesneau, Andreas Kaufer, Asif ud-Doula, Stan Owocki, St\'ephane Mathis, Matteo Cantiello, and Ehsan Moravveji for all of their helpful comments, information, and discussions that went into this work. 

\bibliography{bib_dat.bib}{200}
\label{lastpage}

\end{document}